\documentclass[%
 reprint,
superscriptaddress,
 amsmath,amssymb,
 aps,
twocolumn]{revtex4-2}

\usepackage{latexsym}
\usepackage{graphicx}
\usepackage{verbatim}
\usepackage{multirow}
\usepackage [english]{babel}
\usepackage [autostyle, english = american]{csquotes}
\MakeOuterQuote{"}
\usepackage{amsmath}
\usepackage{mathrsfs}
\usepackage{float}
\usepackage{bm}
\usepackage[usenames, dvipsnames]{color}
\usepackage{tabularx}

\usepackage{mathtools}
\usepackage{slashed}
\usepackage{physics}	% installed packages for typing maths and physics
\usepackage{graphicx}   % need for figures
\usepackage{epstopdf}
\usepackage{hyperref}   % use for hypertext links, including those to external documents and URLs
\usepackage{bbold}
\usepackage{wasysym}
\usepackage{amsmath}
\usepackage{bm}
\usepackage{pdfsync}
\newcommand{\rr}[1]{\mathrm{#1}}
\newcommand{\bu}{\bm{u}}

\newcommand{\misfitr}[3]{ V^{\mathrm{GSFE}}_{#1}(\bm{B}^{#2 \rightarrow #3})}
\renewcommand{\vec}[1]{\bm{#1}}

\begin{document}

% \title{Lattice Structure and Magic Angles of p,q TTG }
\title{Multi-moir\'e trilayer graphene: lattice relaxation, electronic structure, and magic angles}
\author{Charles Yang} 
\affiliation{Department of Physics, Stanford University, Stanford, CA 94305, USA}

\author{Julian May-Mann} 
\affiliation{Department of Physics, Stanford University, Stanford, CA 94305, USA}

\author{Ziyan Zhu} 
\affiliation{Stanford Institute for Materials and Energy Sciences,  SLAC National Accelerator Laboratory,  Menlo Park, CA 94025, USA}

\author{Trithep Devakul} 
\email{tdevakul@stanford.edu}
\affiliation{Department of Physics, Stanford University, Stanford, CA 94305, USA}

\begin{abstract}
We systematically explore the structural and electronic properties of twisted trilayer graphene systems. 
In general, these systems are characterized by two twist angles, which lead to two incommensurate moiré periods. 
We show that lattice relaxation results in the formation of domains of periodic single-moir\'e structures only for twist angles close to the simplest fractions.
For the majority of other twist angles, the incommensurate moir\'e periods lead to a quasicrystalline structure.
We identify experimentally relevant magic angles at which the electronic density of states is sharply peaked and strongly correlated physics is most likely to be realized.

% At small angles, the incommensurability of the moiré periods is small, and is only manifest over the supermoiré scale $\sim a_0/|\theta_{12}\theta_{23}|$. We show that at the moiré scale, the system is well described by two commensurate moiré lattices, with a relative shift that varies on the supermoiré scale. The incommensurate supermoiré structure is heavily affected by lattice relaxation effects. For small $\text{p},\text{q}$, the super-moiré structure consists of large regions with a single moiré period, separated by sharp domain walls. As $\text{p}$ and $\text{q}$ increase, the size of the commensurate regions shrink, and the domain walls broaden. We calculate the local density of states DOS at the moiré scale. Using this, and the relaxed lattice structure, we identify experimentally relevant magic angles where the global DOS is maximized. 
\end{abstract}
\maketitle
\section{Introduction}

The electronic density of states (DOS) serves as a key predictor of strong electronic correlations for a wide range of many-body quantum systems. 
This principle is clearly illustrated by two foundational theories in condensed matter physics.
First, the Stoner criterion, which posits that spontaneous ferromagnetism occurs when the DOS exceeds the inverse interaction energy scale. %\cite{blundell2001magnetism}. %alternatively "occurs when the product of the DOS and the inverse interaction energy scale exceeds 1."
Second, the Bardeen–Cooper–Schrieffer (BCS) theory of superconductivity, where the transition temperature is a monotonically increasing function of the DOS. %\cite{tinkham2004introduction}. 
% The connection between the DOS and correlated physics has only become more concrete in light of modern studies on twisted graphene multilayers. In monolayer graphene, the DOS vanishes at the Fermi energy due to its Dirac-like dispersion and correlated phenomena, like superconductivity, are not observed down to the lowest experimentally accessible temperatures. But with two or more graphene layers, it is possible to tune the DOS at the Fermi energy by twisting adjacent layers and find ``magic angles'' where the DOS is maximized. 
This principle has played out magnificently in twisted bilayer graphene where, following theoretical predictions of a ``magic angle'' at which the DOS exhibits a sharp peak~\cite{bistritzer2011moire,morell2010flat}, experiments revealed a rich interacting phase diagram of correlated insulators, understood to be various forms of flavor ferromagnetism, along with regions of superconductivity\cite{cao2018unconventional, cao2018correlated, yankowitz2019tuning, kerelsky2019maximized, xie2019spectroscopic, jiang2019charge, choi2019electronic, sharpe2019emergent, lu2019superconductors, serlin2020intrinsic, saito2020independent, zondiner2020cascade, wong2020cascade, rozen2021entropic, tschirhart2021imaging, yu2022correlated,tseng2022ahe, zhang2023local}
%\cite{polski2022hierarchy,lin2022spinorbit}
% Famously, the magic angle of bilayer graphene was theoretically determined to be $1.1^\circ$\cite{bistritzer2011moire, morell2010flat}, and subsequent experimental studies of magic angle twisted bilayer (MATBG) devices have revealed a rich interacting phase diagram of correlated insulators, understood to be various forms of flavor ferromagnetism, along with regions of superconductivity.
These discoveries ushered a wave of experimental and theoretical activity in the field of moir\'e materials, now expanded beyond purely graphene, and has led to the experimental realization of a remarkably broad range of exotic strongly correlated states~\cite{andrei2021marvels,mak2022semiconductor}.

%Following pioneering theoretical and experimental work, twisted moiré heterostructures have emerged as highly tunable platforms for realizing strongly correlated quantum phases of matter. By twisting adjacent layers, it is possible to realize magic angles, where the (DOS) at the Fermi-surface is maximized. The single-particle DOS has historically been a key indicator of strongly interacting physics, such as superconductivity, ferromagnetic, and topological order. This expectation has proven to hold for moiré materials. This has been most well appreciated in the case of twisted bilayer graphene, where experiments performed at the theoretically predicted magic angle of magic angle of $\theta \sim 1.1^\circ$, have revealed a rich interacting phase diagram of correlated insulators, understood to be various forms of flavor ferromagnetism, along with regions of superconductivity []. Subsequent research has led to a wave of experimental and theoretical activity in the field of strongly correlated moir\'e materials, now expanded beyond purely graphene. %, and has led to the experimental realization of a broad range of exotic strongly correlated states, culminating in recent observations of the fractional quantum anomalous Hall effect[MoTe2, 5Gr/BN].

While the majority of studies have focused on bilayer systems, going to three or more layers greatly increases the space of possibilities. 
% As is clear from recent observations of the elusive fractional quantum anomalous Hall effect\cite{neupert2011fractional} and composite Fermi liquids\cite{halperin1993theory} in pentalayer graphene/hexagonal boron nitride structures\cite{lu2023fractional}, more layers can lead to phenomena not seen with fewer layers.
The large parameter space of multilayer systems allows for systematic identification of which material properties are necessary for realizing various strongly correlated phenomena, such as superconductivity or orbital ferromagnetism. 
Gaining a better understanding of multilayer systems is therefore paramount to utilizing the full potential of twisted materials.
However, the large and mostly unexplored phase space of multilayer systems also presents a practical barrier to efficient experimental exploration. 
Because of this, it is important to first determine the experimental parameters that are most likely to lead to correlated phenomena in multilayer devices.

We focus our attention on twisted trilayer graphene (TTG) systems, which are characterized by the two twist angles: the angle of rotation between the first and second layers ($\theta_{12}$), and between the second and third layers ($\theta_{23}$)~\cite{morell2013electronic, mora2019flatbands, li2019electronic, zhu2020twisted,tritsaris2020electronic,meng2023commensurate}. 
We will denote specific configurations with the ordered pair $(\theta_{12},\theta_{23})$. To avoid redundancy, we take $|\theta_{12}|<|\theta_{23}|$ throughout this work, as $\theta_{12}\rightarrow \theta_{23}$ when the graphene stack is inverted. 
As we will describe, this two-dimensional parameter space contains many distinct classes of systems, with present experiments showing a wide array of different physical phenomena. 
% Nevertheless, as we discuss below for several illustrative examples, the correlated physics of TTG characteristically occurs at twist angles where the DOS is enhanced, even when the system does not possess a well-defined band structure. 
%We therefore arrive at an answer to our previous question: in the search for new strongly correlated phases of in TTG devices, experiments should consider angles where the DOS is peaked. \textbf{JMM: Need a better punchline here}. 
% In this work, we comprehensively analyze this class of systems and identify twist angle combinations that are likely to host strongly correlated physics, guiding the way for future experimental exploration.

The most well studied TTG systems are ``single-moir\'e'' structures, characterized by a single moir\'e periodicity shared among all layers.
% Like twisted bilayer graphene, trilayers typically also have magic angles characterized by a peak in the DOS indicating the formation of flat bands.
% The flat bands at these magic angles give rise to rich interacting physics, notably
This is the case for the mirror symmetric $(\theta,-\theta)$ TTG~\cite{khalaf2019magic,park2022robust, zhang2021ascendance}.
Near the magic angle of $1.56^\circ$, the central bands become flat, and $\sim (1.6^\circ,-1.6^\circ)$-TTG has been found to exhibit strong correlations and superconductivity with reentrant behavior and possible Pauli limit violations ~\cite{park2021tunable, hao2021electric, cao2021pauli, liu2022isospin,  kim2022evidence, talantsev2022compliance, shen2023dirac}. 
A single-moiré structure is also present in $(\theta,0)$-TTG (i.e. twisted monolayer-bilayer graphene), where topology and ferromagnetism have been observed at twist angles $\sim (1.1^\circ,0)$\cite{polshyn2020electrical,chen2021electrically,He_2021,Xu_2021,Polshyn_2021,zhang2023local}.  %, where a layer bias can cause the system to mimic MATBG or twisted double bilayer graphene. 
Additionally, if $\theta_{23}$ is significantly larger than $\theta_{12}$, the first layer effectively electronically decouples from other layers~\cite{sanchez2012quantum, sanchez2017helical}, such that the only relevant moiré period is set by $\theta_{12}$.
In Ref.~\onlinecite{hoke2023uncovering}, it was shown that $\sim (1.1^\circ,-5.5^\circ)$-TTG behaves like decoupled monolayer graphene and MATBG, and the flavor symmetry breaking commonly found in MATBG was observed.

Generally, TTG realizes a ``multi-moir\'e'' structure in which the moir\'e superlattices formed by $\theta_{12}$ and $\theta_{23}$ have different periods and/or orientations.
Despite not having a single well-defined moiré periodicity, experimental studies have still found evidence of strong electronic interactions in multi-moir\'e systems, indicating that strict moiré periodicity is not a prerequisite for correlated physics. 
% For example, Ref.~\onlinecite{zhang2021correlated} demonstrated samples with supermoir\'e lattices showing correlated insulating states and zero-resistance states.
In Ref.~\onlinecite{uri2023superconductivity}, strong interactions in the form of a ``Dirac revival'', indicative of flavor ferromagnetism, and regions of superconductivity were observed in $(1.4^\circ,-1.9^\circ)$-TTG, termed a ``moir\'e quasicrystal''. 
% While the MQC does not possess Bloch bands, as the top and bottom moiré superlattices form an incommensurate quasiperiodic structure, the DOS is still peaked at this twist angle. %Despite not possessing Bloch bands, theoretical calculations have shown that the correlated phases of MQC are still accompanied by a peak in the single particle DOS at this twist angle, similar to what occurs in single moiré systems with flat bands.
%Cite Ke Wang moire of moire
Very recently, orbital ferromagnetism and an anomalous Hall effect were observed in ``helical trilayer graphene'',  $\sim (1.8^\circ,1.8^\circ)$-TTG~\cite{xia2023helical}. Although the unrelaxed lattice structure of $(\theta,\theta)$-TTG exhibits a multi-moiré pattern, theoretical studies have shown that it relaxes into large domains (several hundreds of nanometers across) of a single-moir\'e structure separated by a triangular network of domain walls\cite{devakul2023magic,kwan2023strong,guerci2023chern,nakatsuji2023multi}. 
The local band structure for these domains contains flat bands at a magic angle of $(1.85^\circ,1.85^\circ)$. 

The TTG systems listed above are strikingly dissimilar in many ways:  $(1.6^\circ,-1.6^\circ)$-TTG and $(1.1^\circ , 0)$-TTG are periodic, $(1.1^\circ,-5.5^\circ)$-TTG is effectively electronically periodic, $(1.8^\circ,1.8^\circ)$-TTG relaxes into periodic domains, and the $(1.4^\circ,-1.9^\circ)$-TTG is quasiperiodic. 
These systems also exhibit different physical phenomena. 
But in all cases, strongly correlated phases have been observed.  
A common thread between all these systems is a strongly peaked DOS appearing near a magic angle.
We use this principle to comprehensively analyze an extensive class of multi-moiré TTG systems and identify experimentally relevant magic angles where strongly correlated phases are likely to be observed, guiding the way for future experimental exploration.

\begin{table}
    \centering
    \bgroup
    % \large
    \def\arraystretch{1.5}
    \begin{tabular}{|c|c|c||c|c|c|}
    \hline
         $\theta_{12}/\theta_{23}$ & \;($\theta_{12}$,\;$\theta_{23}$)\;& \;Exp\;& $\theta_{12}/\theta_{23}$ & ($\theta_{12}$,\;$\theta_{23}$) &  Exp\\
         \hline\hline
         -1/1&$(1.56^\circ,-1.56^\circ)$&\cite{park2022robust, zhang2021ascendance,park2021tunable, hao2021electric, cao2021pauli, liu2022isospin,  kim2022evidence}
         &
         1/1&$(1.85^\circ,1.85^\circ)$&\cite{xia2023helical}
         \\
         -3/4&$(1.41^\circ,-1.88^\circ)$&\cite{uri2023superconductivity}
         &
         3/4&$(1.61^\circ,2.14^\circ)$&---
         \\
         -2/3&$(1.32^\circ,-1.99^\circ)$&---
         &
         2/3&$(1.52^\circ,2.28^\circ)$&---
         \\
         -1/2&$(1.22^\circ,-2.44^\circ)$&---
         &
         1/2&$(1.31^\circ,2.61^\circ)$&---
         \\
         -1/3&$(1.16^\circ,-3.48^\circ)$&---
         &
         1/3&$(1.20^\circ,3.61^\circ)$&---
         \\
         -1/4&$(1.13^\circ,-4.52^\circ)$&$\sim$ \cite{hoke2023uncovering}
         &
         1/4&$(1.16^\circ,\,4.64^\circ)$&---
         \\
         \hline
    \end{tabular}
    \egroup
    \caption{First magic angles for TTG for simple angle ratios, and experiments where strongly correlated physics have been observed. The DOS is peaked within a range of $\sim \pm .05 ^\circ$ of the first magic angle. }\label{tab:MAsum}
\end{table}

\begin{figure}
    \centering
    \includegraphics[width=.45\textwidth]{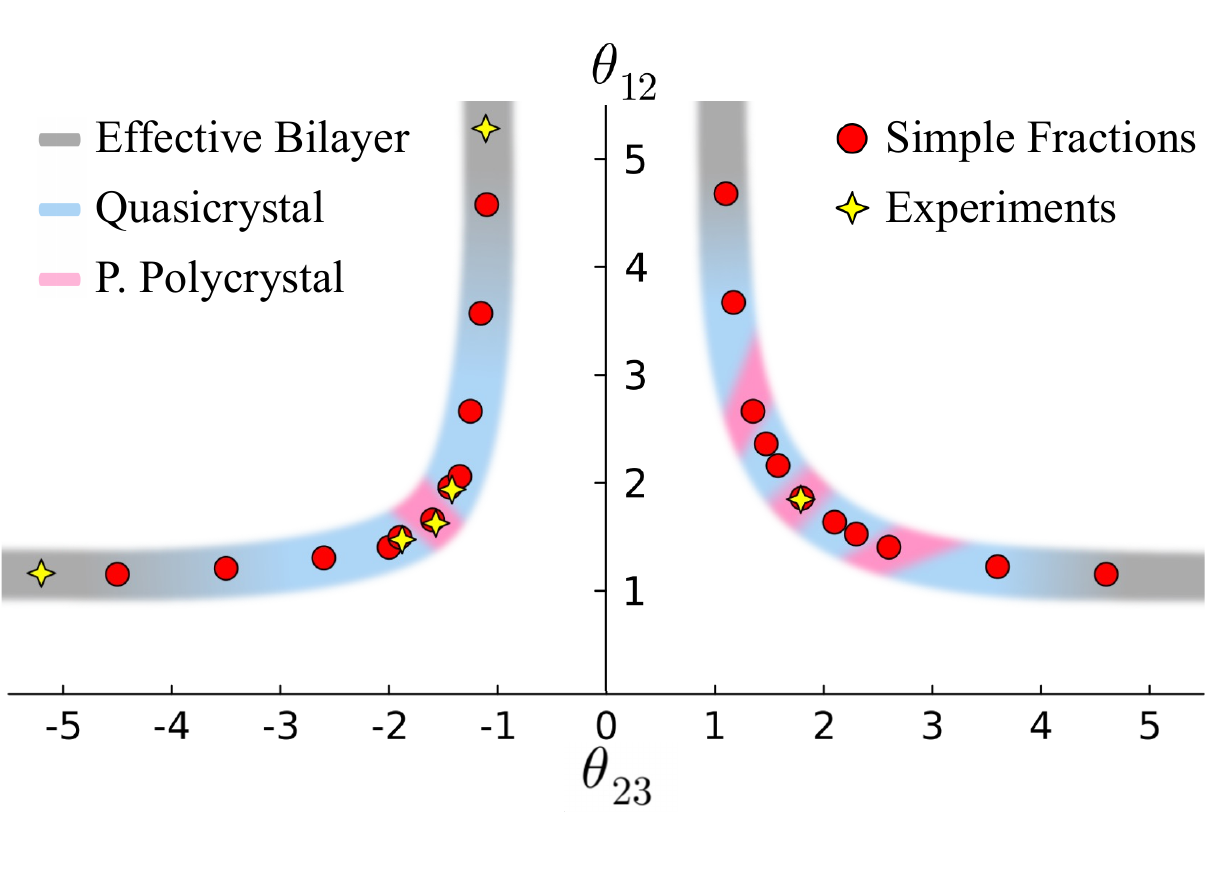}
    \caption[]{The TTG magic angles found in this work (red), where $\theta_{12}/\theta_{23} = \text{p}/\text{q}$ is a simple fraction. The simple fraction magic angles are connected by a "magic line". For moderate values of $\theta_{12}$ and $\theta_{23}$ the system is either a periodic polycrystal (pink) or quasicrystal (blue). For large $\theta_{12}$ or $\theta_{23}$, (gray) one of the layers decouples, and the other two layers form an effective magic angle bilayer. Stars indicate twist angles where correlated phenomena have been observed in experiment: $(1.6^\circ,-1.6^\circ)$\cite{park2021tunable, hao2021electric, cao2021pauli, liu2022isospin,  kim2022evidence, shen2023dirac}, $(1.4^\circ,-1.9^\circ)$\cite{uri2023superconductivity}, $(1.1^\circ,\sim -5.5^\circ)$\cite{hoke2023uncovering}, and $(1.8^\circ,1.8^\circ)$\cite{xia2023helical}. Data is symmetrized with respect to exchange of the first and third layers.  } \label{fig:TTG_Datapoints}
\end{figure}

We first explore the effect of lattice relaxation, which plays a crucial role in multi-moir\'e TTG systems due to the large supermoir\'e period.
Our key result is that the relaxed structure of multi-moir\'e TTG systems, near the first magic angles, broadly fall into three categories, summarized in Fig.~\ref{fig:TTG_Datapoints}.    
(i) Supermoir\'e scale lattice relaxation results in the formation large domains of an energetically favorable single-moir\'e structure, separated by sharp domain walls.  This structure resembles a polycrystal, but with a periodic arrangement of crystalline domains on the supermoir\'e scale.  We thus refer to this as the ``periodic polycrystalline'' regime, which encompasses systems with twist angles close to $\theta_{12}/\theta_{23}\approx \pm1$, and $\theta_{12}/\theta_{23}\approx1/2$.
(ii) Supermoir\'e lattice relaxation does \emph{not} result in the formation of large single-moir\'e domains.  We refer to this as the ``quasicrystalline'' regime, which encompasses much of the $(\theta_{12},\theta_{23})$ phase space away from the polycrystalline cases above (including MQC).  
(iii) Finally, when $|\theta_{12}|$ is significantly smaller than $|\theta_{23}|$, the third layer effectively decouples electronically, and the physics reduces to that of TBG.
We refer to this as the effectively ``bilayer'' regime, which occurs for $|\theta_{12}/\theta_{23}|\lesssim 1/4$.

We further explore the electronic properties of multi-moir\'e TTG.  
When $\theta_{12}/\theta_{23}\approx \text{p}/\text{q}$ is close to a simple fraction, the local moir\'e-scale electronic properties can be well described by a generalized Bistritzer-MacDonald continuum model\cite{khalaf2019magic,mora2019flatbands,mao2023supermoire,popov2023magic,popov2023magic2,devakul2023magic}, parameterized by a shift vector $\mathbf{d}$ between the two moir\'e superlattices.  
We explore the twist angle and $\mathbf{d}$ dependence of the DOS which, combined with the relaxed structure, allow us to identify magic angles at which the DOS is maximized, summarized in Table~\ref{tab:MAsum}. 
We remark that, for systems that have already been explored experimentally, our magic angles agree quantitatively with the experimental angles at which strongly correlated states are found.

This paper is structured as follows: In Sec.~\ref{sec:Lattice_Relaxation}, we present lattice relaxation calculations for multi-moir\'e TTG and discuss how the relaxed lattice forms either a periodic polycrystal or a quasicrystal. In Sec.~\ref{sec:magicangles}, we analyze the electronic structure when the twist angles form a simple fraction, and determine magic angles where the DOS is peaked. We discuss the physical consequences and experimental relevance of our results in Sec.~\ref{sec:disc}. We also provide several appendices that contain technical details of our results. 
\section{Supermoir\'e Lattice Relaxation}\label{sec:Lattice_Relaxation}

Multi-moiré TTG systems are characterized by two twist angles, $\theta_{12}$ and $\theta_{23}$. Explicitly, if we label the twist of layer $l$ relative to some fixed orientation as $\theta_l$, then $\theta_{12} = \theta_1-\theta_2$ and $\theta_{23} = \theta_2-\theta_3$. This leads to two subclasses of multi-moiré TTG systems: those with alternating twist angles, $\theta_{12}/\theta_{23} < 0$, and those with helical twist angles, $\theta_{12}/\theta_{23} > 0$.
% \footnote{Note that ($\theta,\theta$)-HTG has helical twists, but all TTG systems with helical twists are not HTG systems}. 
In both cases, TTG is characterized by several length scales: the graphene lattice constant $a_0= 2.46\,\text{\AA}$, and the moiré scales $ a_{12}\approx a_0/|\theta_{12}|$ and $a_{23}\approx a_0/|\theta_{23}|$.
When $\theta_{12}/\theta_{23}=\text{p}/\text{q}$ is a simple fraction, then there is an additional supermoir\'e lengthscale $a_{sm} \approx 2a_0/|(\text{p}+\text{q})\theta_{12}\theta_{23}|$. 
In the small angle limit, the graphene lattice constant is much smaller than the moiré scale, which in turn is much smaller than the supermoiré scale. Due to this decoupling of length scales, the local structure of TTG at the moiré scale can be well approximated by two commensurate moiré lattices with a relative shift $\mathbf{d}$ between them. 
The global structure can then be restored by allowing $\mathbf{d}$ to vary slowly in space on the supermoir\'e scale.
As we will elaborate, lattice relaxation on the supermoir\'e scale effectively controls how $\mathbf{d}$ varies in space.

To see why multi-moiré TTG can be modeled as locally commensurate, let us first examine the rigidly twisted structure.
Let $\mathbf{G}_l=\frac{4\pi}{\sqrt{3}a_0}\mathbf{R}(\theta_l)(\sqrt{3}/2, -1/2; 0, 1)^T$ be the matrix with columns given by the atomic reciprocal lattice vectors of layer $l$, where $\mathbf{R}(\theta)=(\cos\theta, -\sin\theta; \sin\theta,\cos\theta)$ is a rotation matrix, and $\theta_l$ is the twist angle of layer $l$.
The moir\'e reciprocal lattice vectors for layers $l$ and $l'$ are then given by $\mathbf{G}_{ll'}=\mathbf{G}_l-\mathbf{G}_{l'}$, and the corresponding real-space moir\'e unit vectors is given by the columns of $\mathbf{A}_{ll'}=2\pi\mathbf{G}_{ll'}^{-T}$.  
In general, $\mathbf{G}_{12}$ and $\mathbf{G}_{23}$ are unrelated to each other, and thus lead to an incommensurate structure.
% In special cases, however, a commensurate structure is possible: if there exists a set of integers such that $\mathbf{A}_{12}(n_1,n_2)=\mathbf{A}_{23}(n_3,n_4)\equiv \mathbf{A}$, where not all $n_i=0$, then $\mathbf{A}$ forms an enlarged commensurate unit cell containing an integer multiple of both $A_{12}$ and $A_{23}$.
However, when $\theta_{12}/\theta_{23}=\text{p}/\text{q}$ (excluding the special case of $\text{p}/\text{q}=-1$), 
$||\frac{1}{\text{p}}\mathbf{G}_{12} - \frac{1}{\text{q}}\mathbf{G}_{23}||=\mathcal{O}(\theta^2)$ vanishes at first order in $\theta$.  
This signifies the proximity to a commensurate structure in which $\text{p}\mathbf{A}_{12}\approx \text{q}\mathbf{A}_{23}$.  

\begin{figure}
    \centering
    \includegraphics[width=.45\textwidth]{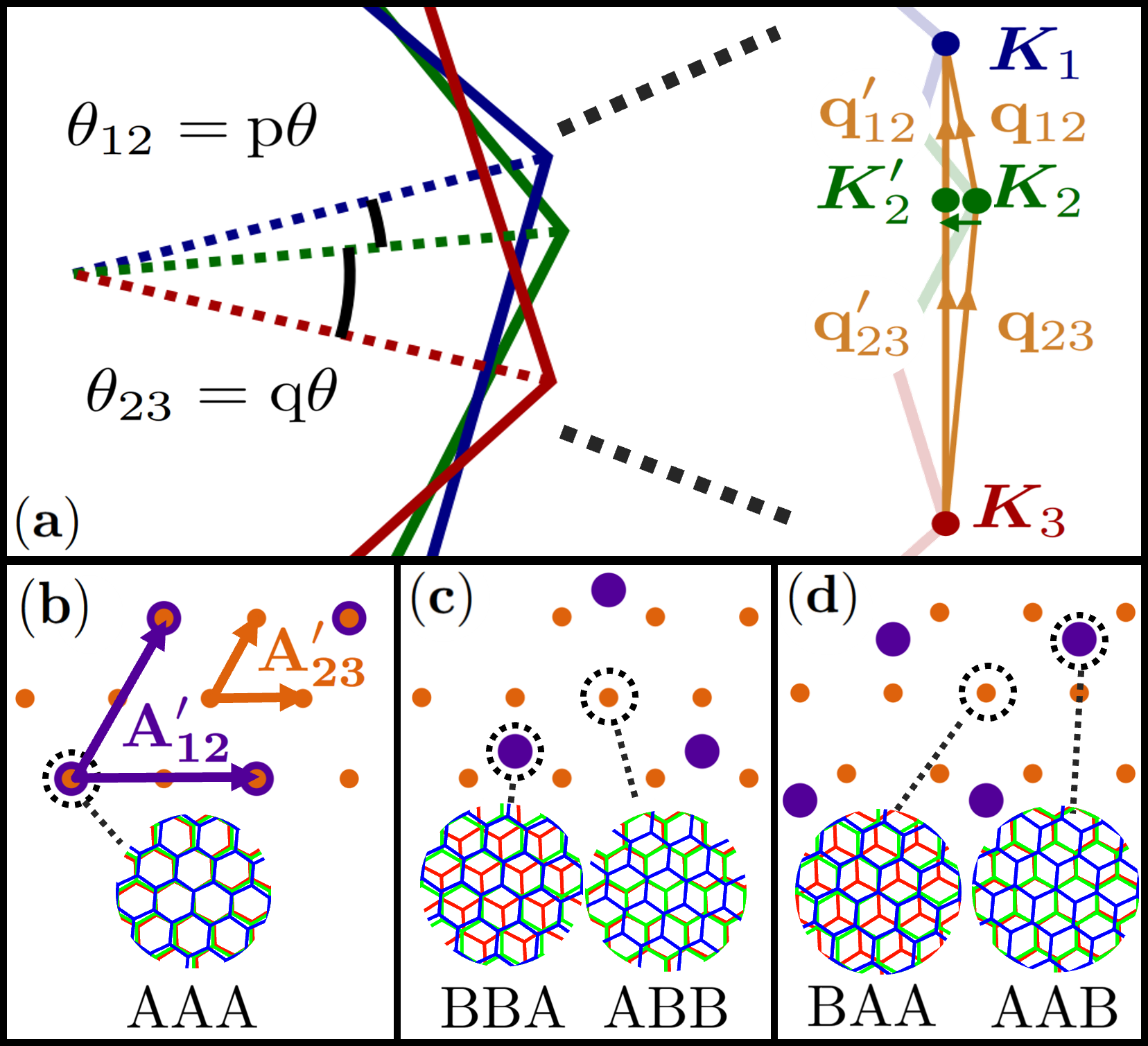}
    \caption[]{(\textbf{a}, left) The Brillouin zones of the three graphene layers. (\textbf{a}, right) A slight biaxial dilation and twist of the second layer, $\bm K_2\rightarrow \bm K'_2$, leads to a configuration where  $\bm K_1 - \bm K'_2 = \mathbf{q}'_{12}$ and  $\bm K'_2 - \bm K_3 = \mathbf{q}'_{23}$ are exactly commensurate: $\mathbf{q}'_{12} = \frac{\text{p}}{\text{q}} \mathbf{q}'_{23}$.  Here, we illustrate the $\text{p}/\text{q}=1/2$ case. (\textbf{b}) The commensurate structure at $\mathbf d = 0$, with the AA stacking regions of layers 1 and 2 in purple, and layers 2 and 3 in orange. The local atomic stacking is illustrated at certain high symmetry points. (\textbf{c},\textbf{d}) Same as (\textbf{b}) but with a shift of $\mathbf d = +\bm \delta, -\bm\delta$ respectively. 
    } \label{fig:distorion}
\end{figure}

This commensurate structure can be made exact by a slight deformation of the layers.  
By applying a slight biaxial dilation and twist to the second layer, $\mathbf{G}_2\rightarrow\mathbf{G}_2^\prime=\lambda\mathbf{R}(\delta\theta)$, with $\lambda\approx 1$ and $\delta\theta\approx 0$, we can ensure that the resulting moir\'e reciprocal lattice vectors satisfy the commensurability condition $\frac{1}{\text{p}}\mathbf{G}_{12}^\prime=\frac{1}{\text{q}}\mathbf{G}_{23}^\prime\equiv\mathbf{G}^\prime$ exactly.
The commensurate unit cell is given by $\mathbf{A}^\prime=\text{p}\mathbf{A}_{12}^\prime=\text{q}\mathbf{A}_{23}^\prime$.
Here, primes indicate the structure with the deformed $\mathbf{G}_2^\prime$ in place of the original $\mathbf{G}_2$.
This procedure is illustrated in Fig.~\ref{fig:distorion} in terms of the $K$ points, $\mathbf{K}_l=\mathbf{G}_l(2/3,1/3)^T$.
In the original structure, the vectors $\mathbf{q}_{12}$ and $\mathbf{q}_{23}$, where $\mathbf{q}_{ll^\prime}=\mathbf{K}_l-\mathbf{K}_{l^\prime}$, are incommensurate.  
However, when $\theta_{12}/\theta_{23}\approx \text{p}/\text{q}$, a slight deformation of the second layer $\mathbf{K}_2\rightarrow\mathbf{K}_2^\prime$ results in a commensurate arrangement in which $\mathbf{q}'_{12} = \mathbf{K}_1 - \mathbf{K}'_2=\frac{\text{p}}{\text{q}}\mathbf{q}'_{23} = \frac{\text{p}}{\text{q}}(\mathbf{K}'_2-\mathbf{K}_3)$.

We remark that, at this point, the uniform biaxial heterostrain and twist is simply a theoretical tool to construct a commensurate periodic approximant.
We do not always expect this kind of deformation to occur in a real system.
% is unlikely to occur globally in a real system since it involves a redistribution of mass from the original structure.
Nevertheless, the commensurate structure is a good approximation for the local physics at the moir\'e scale, as the differences only become apparent over distances $\sim |\bm{K}'_2 - \bm{K}_2|^{-1}$. 
% The moiré scale physics around a given point of the full multi-moir\'e TTG system can therefore be understood in terms of two commensurate moiré lattices. 
This local commensurate description is characterized by the commensurate moiré reciprocal lattice vectors, $\mathbf{G}'_{ll'}$, and the relative displacement between the two moiré lattices $\mathbf{d}$. 
Explicitly, if we fix some reference point of a given trilayer lattice structure, and define $\mathbf{d}_t$ as the distance from that point to a nearby AA region of layers 1 and 2, and $\mathbf{d}_b$ as the distance to a nearby AA region layers 2 and 3, then we can define
\begin{equation}
    \mathbf{d} = |\text{p}\text{q}| (\mathbf{d}_t -\mathbf{d}_b)\mod(\mathbf{A}')
\end{equation} 
which respects the periodicity of the commensurate structure, and is independent of which nearby AA regions are chosen. 
Clearly, when $\mathbf{d} = 0$, each unit cell of the commensurate structure will contain an AAA stacking. When $\mathbf{d} = \pm \bm{\delta} \equiv \pm \mathbf{A}^\prime (1/3,1/3)^T$
% $(\bm{A}'_{2} - \bm{A}'_{1})/3$
the unit cell will instead contain at least one of an AAB, ABB, BAA, or BBA stacking region.
% Similar to Refs.~\onlinecite{jung2014ab} and~\onlinecite{lei2021mirror}, we assume that different regions are characterized by the same moiré lattices, but with different relative displacements. 

The value of $\mathbf{d}$ at a point $\mathbf{r}$ can also be directly related to local atomic stacking configurations. 
Suppose that, at the reference point $\mathbf{r}$, the local atomic positions in layer $l$ appears shifted by $\mathbf{s}_l$ relative to that of the origin.  Then, the expression for $\mathbf{d}$ is given by (see Appendix for derivation)
\begin{equation}
\mathbf{d}(\mathbf{s}_1,\mathbf{s}_2,\mathbf{s}_3)=\mathbf{A}^\prime\left[q\mathbf{A}_1^{-1} \mathbf{s}_1  - (q+p)\mathbf{A}_2^{-1} \mathbf{s}_2 + p\mathbf{A}_3^{-1}\mathbf{s}_3\right] .
\label{eq:d_eq}
\end{equation}
With this in mind, we now consider the fully relaxed lattice structure of multi-moir\'e TTG.

We model lattice relaxation utilizing the configuration space method developed in \citet{zhu2020modeling}. 
To obtain the relaxation pattern, we minimize the total energy as a function of the relaxation displacement vectors in the local configuration space (see Appendix~\ref{sec:relax} for details)~\cite{cazeaux2020energy}. The total energy is a sum of intralayer and interlayer energies.
The intralayer energy is the elastic energy cost due to the surface deformation, and the interlayer energy is the energy due to the layer misfit, which is parametrized by the Generalized Stacking Fault energy (GSFE)~\cite{kaxiras1993free,zhou2015vdw}. 
Both of these terms can be obtained with first-principles Density Functional Theory~\cite{carr2018relaxation,zhu2020modeling}. 

We compute the local shift field $\mathbf{u}_l(\mathbf{r})$ for the fully relaxed structure at a variety of twist angles.  
At a point $\mathbf{r}$, the local atomic positions are shifted by $\mathbf{s}_l(\mathbf{r})=-\mathbf{r}+\mathbf{u}_l(\mathbf{r})$.  
Plugging this into Eq.~\eqref{eq:d_eq} gives an expression for the $\mathbf{d}(\mathbf{r})$ in the relaxed structure as a function of position.
The quantity $\mathbf{d}(\mathbf{r})$ should be understood as the following: if we look at the local atomic structure at $\mathbf{r}$ and extend it infinitely into the commensurate structure $\mathbf{A}^\prime$, then this structure would be characterized by the constant shift vector $\mathbf{d}(\mathbf{r})$ between the two moir\'e patterns.
In the unrelaxed structure, $\mathbf{s}_l(\mathbf{r})=-\mathbf{r}$, and $\mathbf{d}(\mathbf{r})$ varies linearly with $\mathbf{r}$, and is periodic in $\mathbf{r}$ with the supermoir\'e periodicity $a_\mathrm{sm}$.
In the presence of lattice relaxation, we expect that $\mathbf{d}(\mathbf{r})$ will exhibit some moir\'e-scale fluctuations due to lattice relaxation at the moir\'e scale, but more importantly, it should capture the long wavelength supermoir\'e scale relaxation.
In particular, if the system relaxes into large domains of a commensurate structure, $\mathbf{d}(\mathbf{r})$ should remain roughly constant throughout the entire domain (up to moir\'e-scale fluctuations).

\begin{figure}
    \centering
    \includegraphics[width=1\linewidth]{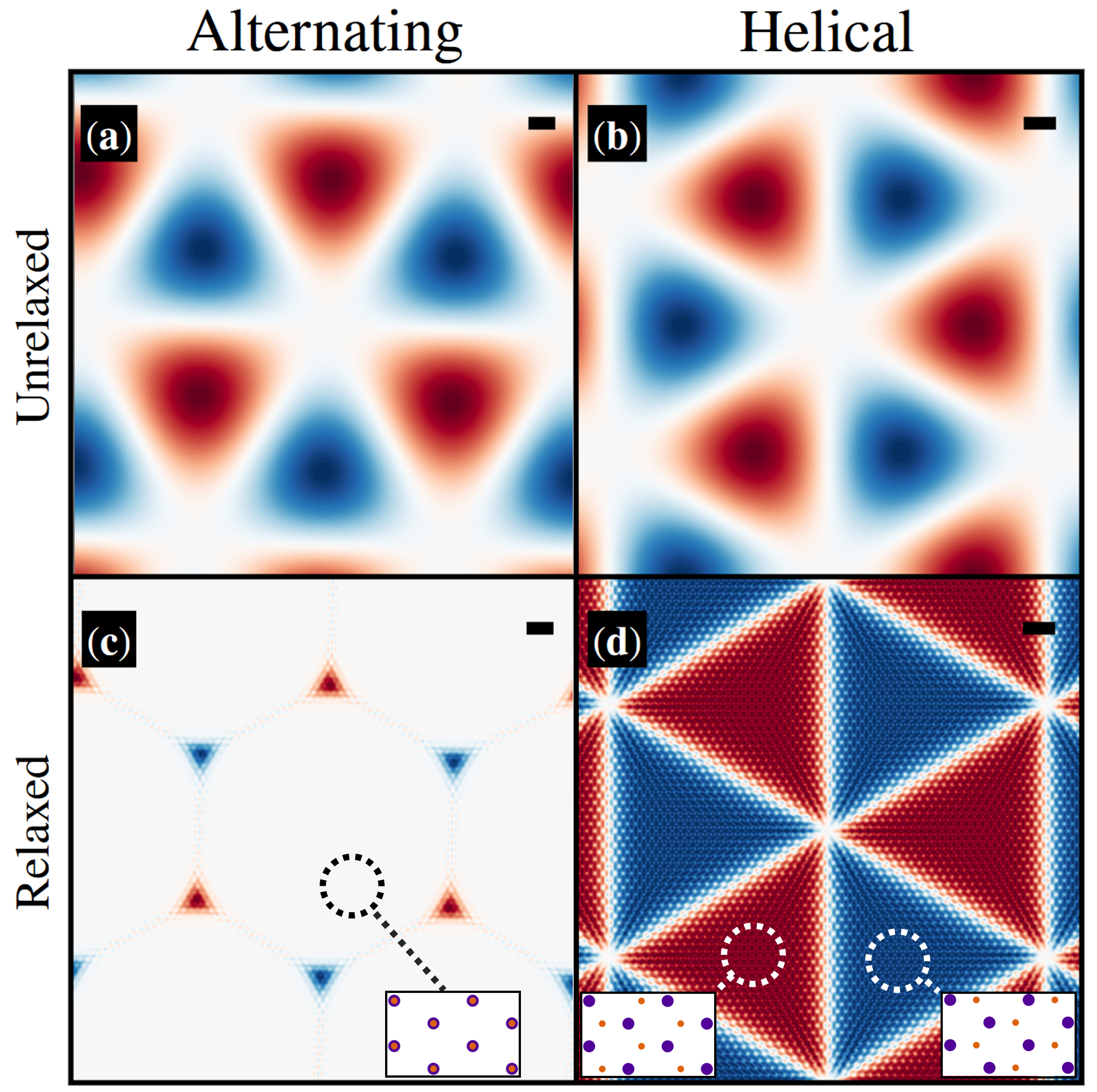}
    \caption{ The value of the function, $W(\mathbf{d}(\mathbf{r}))$, for twist angles with $|\theta_{12}/\theta_{23}|\approx 1$. 
    The unrelaxed structure is shown for \textbf{(a)} almost mirror symmetric $(1.60^\circ,-1.65^\circ)$-TTG and \textbf{(b)} $(1.85^\circ,1.85^\circ)$-TTG. $\mathbf{d} = 0$ in the white regions, and $\mathbf{d} = \pm \bm{\delta}$ in the red and blue regions, respectively. The relaxed lattice structures is shown for \textbf{(c)} $(1.60^\circ,-1.65^\circ)$-TTG, which forms large domains where $\mathbf{d} = 0$, and \textbf{(d)} $(1.85^\circ,1.85^\circ)$-TTG,  forms large domains where $\mathbf{d} = \pm \bm{\delta}$. 
    Insets show the relative shift of the AA stacking regions of each layer pair, as in Fig.~\ref{fig:distorion}.  The scale bar is 30nm.}
    \label{fig:relax11}
\end{figure}

\begin{figure*}
    \centering
    \includegraphics[width=1\linewidth]{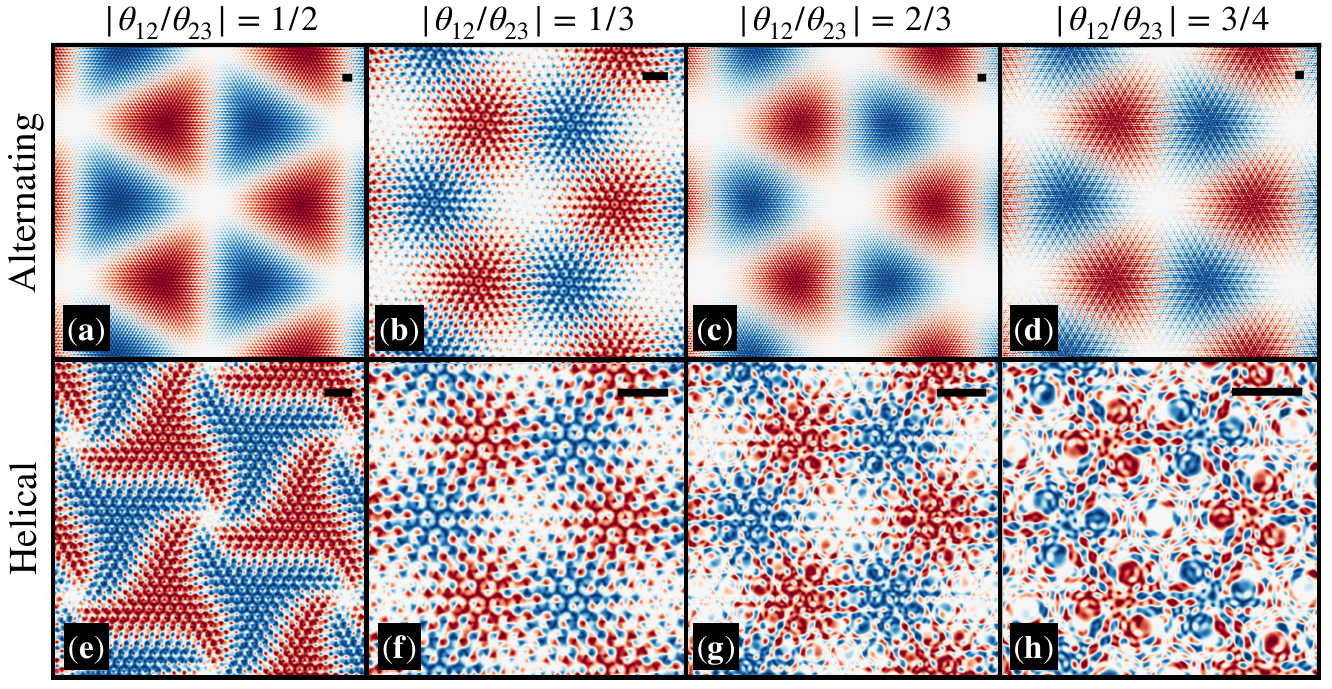}
    \caption{ The function $W(\mathbf{d}(\mathbf{r}))$ is plotted for the relaxed lattice structure for alternating ($\theta_{12}/\theta_{23} < 0$) and helical ($\theta_{12}/\theta_{23} > 0$) TTG near the first magic angles: \textbf{(a)} $(1.30^\circ, -2.60^\circ)$,  \textbf{(b)} $(1.20^\circ, -3.60^\circ)$, \textbf{(c)} $(1.40^\circ, -2.10^\circ)$, \textbf{(d)} $(1.50^\circ, -2.00^\circ)$, \textbf{(e)} $(1.25^\circ, 2.50^\circ)$, \textbf{(f)} $(1.20^\circ, 3.60^\circ)$, \textbf{(g)} $(1.50^\circ, 2.25^\circ)$, and \textbf{(h)} $(1.60^\circ, 2.13^\circ)$. The scalebar represents 30nm.
    Only for helical $\text{p}/q=1/2$ does the structure relax into a polycrystalline structure, with clear blue/red domains forming.
    For the remaining fractions, the long wavelength behavior resembles that of the unrelaxed case (Fig.~\ref{fig:relax11}b).
    }
    \label{fig:relax_general}
\end{figure*}

In Fig.~\ref{fig:relax11}, we show how $\mathbf{d}(\mathbf{r})$ behaves for two illustrative examples: almost mirror-symmetric $(1.60^\circ,-1.65^\circ)$-TTG and $(1.85^\circ,1.85^\circ)$-TTG.  
Previous studies have shown that $(\theta,-\theta)$-TTG relaxes towards a single-moir\'e structure with $\mathbf{d}=0$~\cite{carr2020coexistence,turkel2022orderly}, while $(\theta,\theta)$-TTG relaxes into domains of the single-moir\'e structure with $\mathbf{d}=\pm\bm{\delta}$~\cite{zhu2020modeling,devakul2023magic,nakatsuji2023multi}.
We plot the function $W(\mathbf{d})=\sum_{j=1,2,3}\sin(\mathbf{g}_j^\prime\cdot\mathbf{d})$, where $(\mathbf{g}_1^\prime,\mathbf{g}_2^\prime)=\mathbf{G}^\prime$ and $\mathbf{g}_3^\prime=-\mathbf{g}_1^\prime-\mathbf{g}_2^\prime$.
$W(\mathbf{d})$ is zero when $\mathbf{d}=0$, and approaches its maximum/minimum at $\mathbf{d}=\pm\bm{\delta}$.  
In the unrelaxed structure, Fig.~\ref{fig:relax11}(a,b), $W(\mathbf{d}(\mathbf{r}))$ varies smoothly in space with the supermoir\'e periodicity, as expected.
In the relaxed structure, Fig.~\ref{fig:relax11}(c,d), we find $\mathbf{d}(\mathbf{r})$ relaxes strongly to large hexagonal domains with $\mathbf{d}(\mathbf{r})\approx 0$ for alternating TTG, and large triangular domains with $\mathbf{d}(\mathbf{r})\approx\pm\bm{\delta}$ for helical TTG, consistent with earlier findings.

We now proceed to characterize twist angles at other rational fractions $\theta_{12}/\theta_{23}=\text{p}/\text{q}$.  
For each fraction, we focus on twist angles near the magic angles (determined later in Sec~\ref{sec:magicangles}).  
Our results for the relaxed structure are shown in Fig.~\ref{fig:relax_general}, and should be compared to the unrelaxed structure which is identical to Fig.~\ref{fig:relax11}b up to an overall rescaling.  
We highlight several key observations.

For alternating twist angles $\theta_{12}/\theta_{23}<0$ (except for near $\theta_{12}/\theta_{23}=-1$), we find that, despite the very large supermoir\'e period, the effect of lattice relaxation is weak and $\mathbf{d}(\mathbf{r})$ closely resembles that of the unrelaxed case.
This implies that, for alternating twists with $\theta_{12}/\theta_{23}\neq 1$, the interlayer elastic energy gain from forming a locally commensurate  structure $\mathbf{A}^\prime$ is smaller than the intralayer elastic energy cost required to do so.
These findings are surprising, given how strongly the $\theta_{12}/\theta_{23}\approx-1$ case (Fig.~\ref{fig:relax11}c) relaxes towards the $\mathbf{d}=0$ structure.

For helical twist angles, $\theta_{12}/\theta_{23}>0$, we find that lattice relaxation has a stronger effect.  
For $\theta_{12}/\theta_{23}=1/2$, we find domains of commensurate $\mathbf{d}=\pm\bm{\delta}$ regions, separated by a network of curved domain walls.  
For other fractions $1/3,2/3,$ and $3/4$, we find that relaxation does not result in the formation of large periodic domains, but still leads to visible moir\'e-scale fluctuations.  

Our results demonstrate that lattice relaxation effects are generally more significant for helical twists than alternating twists.
Only for the $\theta_{12}/\theta_{23}=\pm1$, and for the helical $\theta_{12}/\theta_{23}=1/2$, does multi-moir\'e TTG relax into large domains of a single-moir\'e structure.
This structure, consisting of large domains of roughly constant $\mathbf{d}$ separated by sharp domain boundaries, is reminiscent of a polycrystal, except that different grains have a periodic structure instead of being randomly arranged.  
We thus refer to these structures as periodic polycrystals~\footnote{We thank Aviram Uri for inspiration for this name}.
For all other angles, $\mathbf{d}$ varies smoothly in position (up to the moir\'e-scale fluctuations) similar to the unrelaxed case.
This structure is characterized by a set of two incommensurate reciprocal lattice vectors, $\mathbf{G}_{12}$ and $\mathbf{G}_{23}$,
and is therefore a moir\'e quasicrystal~\cite{uri2023superconductivity}.

We remark that our conclusions of the relaxed structure will generally depend on magnitude of twist angle, not just their ratio.  
Our results are valid for twist angles near the first magic angle, which are of most potential interest for strongly correlated physics.
At smaller or larger overall twist angles, qualitatively different relaxation patterns can occur.  
% For instance, Ref~\cite{nakatsuji2023multi} found that $(0.62^\circ,-1.47^\circ)$-TTG relaxed into regions of nearly $\mathbf{d}=0$, $\theta_{12}/\theta_{23}=-1/2$ structure, qualitatively different from our conclusions at the first magic angle.  
% To confirm that the difference is indeed due to the smaller twist angle difference, and not due to differences in modeling, we also show the relaxed structure using our method at smaller twist angles (Appendix).
Finally,  high resolution images of the moir\'e AA stacking regions in the relaxed structure are available in the supplemental material.

\section{Local Hamiltonian and Magic Angles}
\label{sec:magicangles}
\begin{figure}
    \centering
    \includegraphics[width=.48\textwidth]{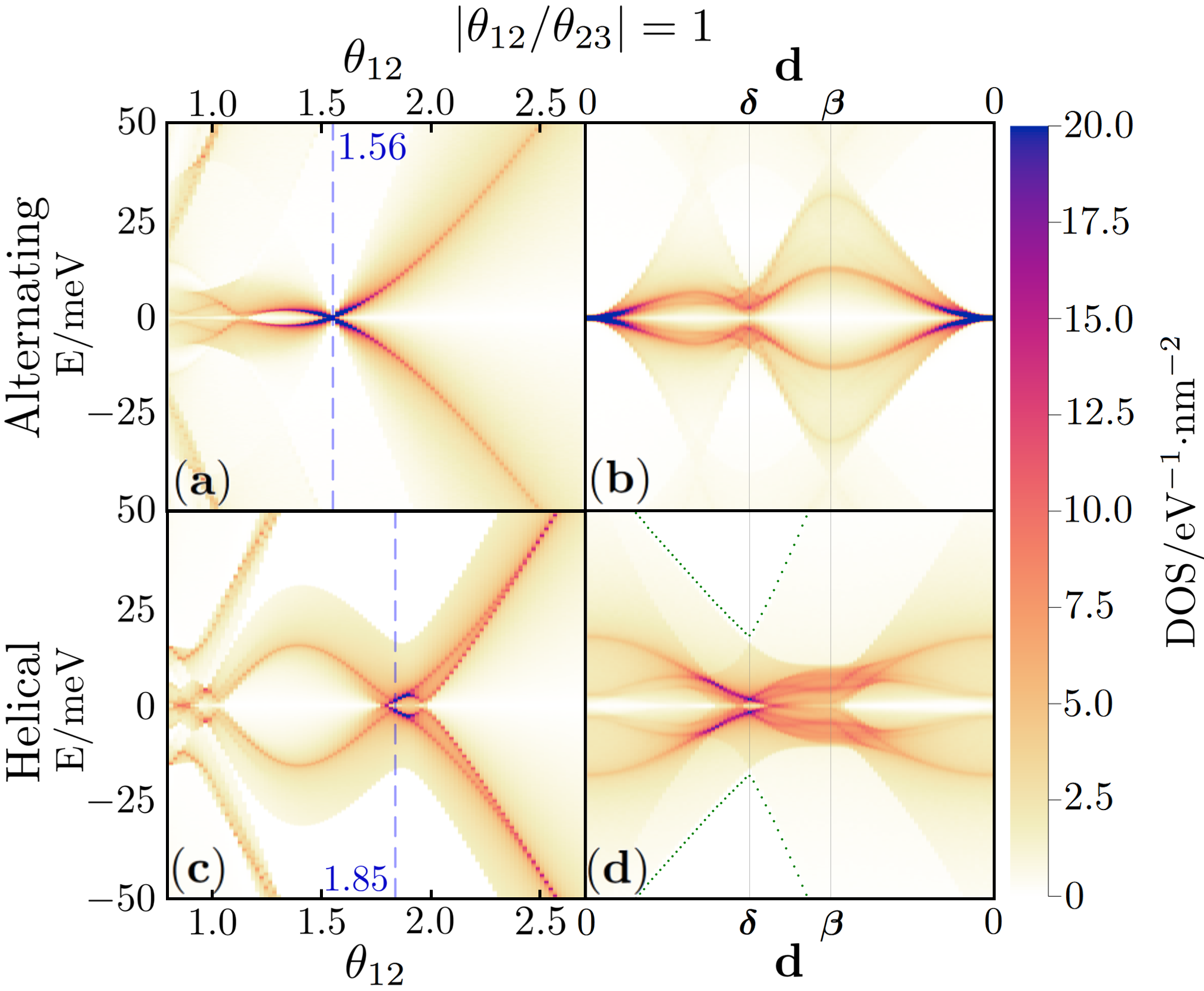}
    \caption[]{(\textbf{a},\textbf{c}) The DOS as a function of twist angle, calculated using Eq.~\eqref{eq:bm} for two known cases. \textbf{(a)} twist angles $\theta_{12} = - \theta_{23}$, and shift $\mathbf{d} = 0$. \textbf{(c)} twist angles $\theta_{12} = \theta_{23}$, and shift $\mathbf{d} = \bm{\delta}$. The magic angles of $1.56^\circ$ and $1.85^\circ$ where the DOS is peaked (dashed line) correspond to magic angles of Refs.~\onlinecite{park2021tunable, hao2021electric, cao2021pauli, liu2022isospin,  kim2022evidence, shen2023dirac} and \onlinecite{xia2023helical} respectively. \textbf{(b},\textbf{d)} The DOS at the magic angle as a function of moir\'e shift, using the path described in Fig.~\ref{fig:DOS_vs_Shift}(\textbf{b}).} \label{fig:DOS_vs_TwistAngle_HTG}
\end{figure}

\begin{figure*}
    \centering
    \includegraphics[width=\textwidth]{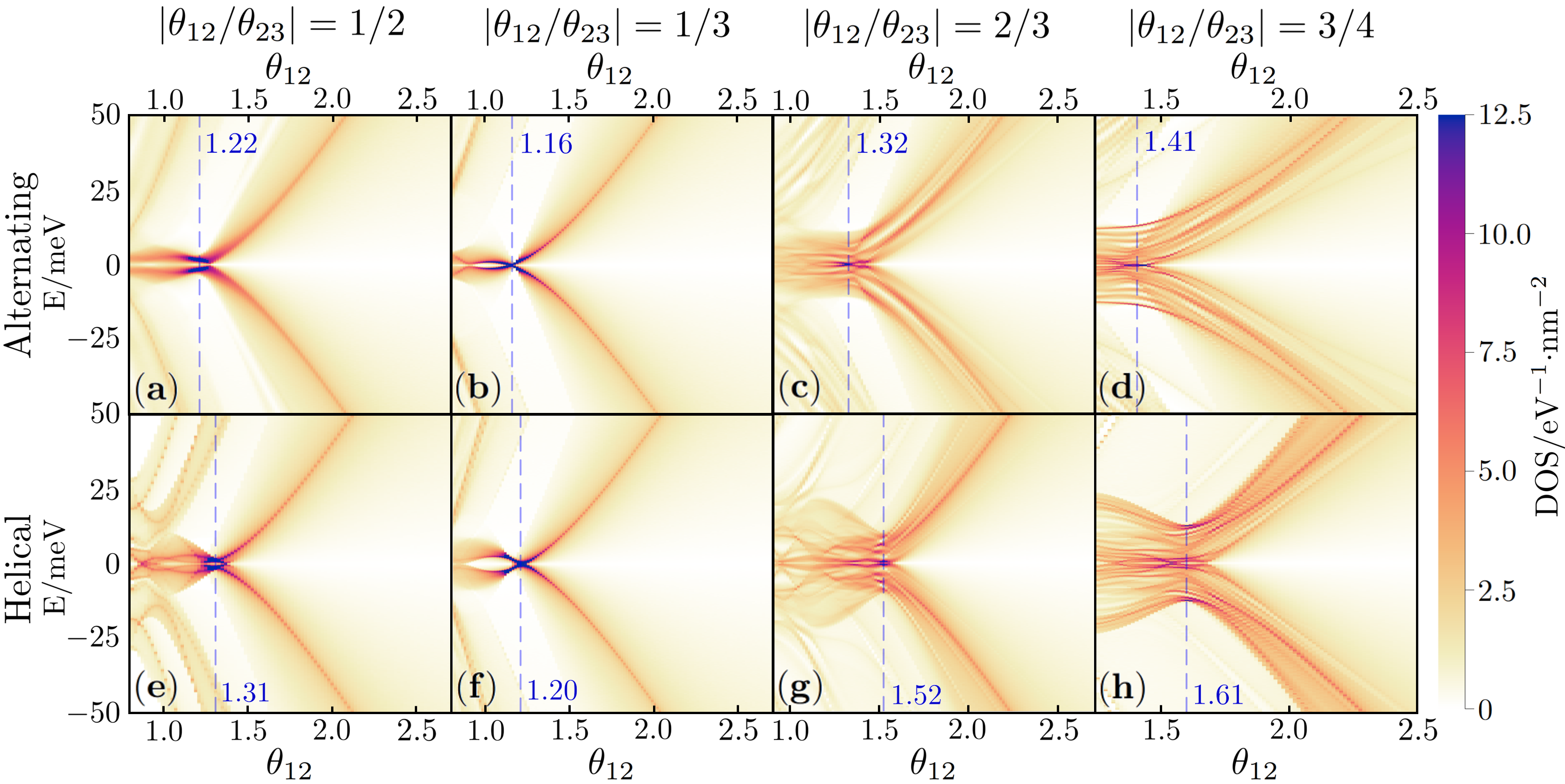}
    \caption[]{The local DOS calculated as a function of twist angle, $\theta_{12} = \frac{\text{p}}{\text{q}} \theta_{23}$, calculated using Eq.~\eqref{eq:bm}. (\textbf{e}) Helical $1/2$ is a periodic polycrystal, so the dominant shift of $\mathbf d = \bm\delta$ was used. (\textbf{a}-\textbf{d},\textbf{f}-\textbf{h}) For the remaining structures, the DOS is averaged over all shifts. The magic angle where the DOS is peaked is estimated with a dashed blue line. The peaks of the DOS have a width of $\sim .1^\circ$.} \label{fig:DOS_vs_TwistAngle}
\end{figure*}

In this section, we will consider the electronic structure of multi-moiré TTG.
% Here, we consider the local electronic properties at the moiré scale
% , as the electronic wavefunctions are not expected to coherent over super-moiré scales.
Following our discussion in Sec.~\ref{sec:Lattice_Relaxation}, we treat local moiré scale lattice structure of multi-moiré TTG as two commensurate moiré lattices with a relative shift between them. Such a system is described by a generalized Bistritzer-MacDonald model~\cite{foo2023extended,popov2023magic,devakul2023magic,popov2023magic2,guerci2023nature}. The Hamiltonian for the K-valley is
\begin{equation}
    H = \begin{pmatrix}
        -iv_F\boldsymbol\sigma_{\theta_{12}}\cdot\boldsymbol\nabla & T_{12}(\mathbf{r}-\mathbf{d}_t) & 0\\
        T_{12}^\dagger(\mathbf{r}-\mathbf{d}_t) & -iv_F\boldsymbol\sigma\cdot\boldsymbol\nabla & T_{23}(\mathbf r - \mathbf d_b)\\
        0 & T_{23}^\dagger(\mathbf r - \mathbf d_b) & -iv_F\boldsymbol\sigma_{-\theta_{23}}\cdot\boldsymbol\nabla
    \end{pmatrix}\label{eq:bm}
\end{equation}
where $\boldsymbol\sigma_{\theta}=e^{-i\theta\sigma_3/2}(\sigma_1,\sigma_2)e^{i\theta\sigma_3/2}$,
and $\mathbf{d}_{t,d}$ is displacement of the between the top (12) and bottom (23) moir\'e lattice respectively. In practice, the Pauli rotation simply results in a small particle-hole asymmetry, and so we make the approximation $\boldsymbol\sigma_\theta\approx\boldsymbol\sigma$. Since a uniform displacement of both moiré lattices is trivial, the spectrum can be parameterized in terms of the shift $\mathbf{d}= |\text{p}\text{q}|(\mathbf{d}_{t} -\mathbf{d}_{d})$. The shift is odd under $C_{2z}\mathcal{T}$ symmetry $\mathbf{d} \rightarrow -\mathbf{d}$  (the product of two-fold rotations, and time-reversal symmetry).

The tunneling matrices between adjacent layers $l$ and $l'$ are
\begin{subequations}
\begin{align}
T_{ll'}(\vb r) &= w\sum_{n=0}^2 e^{i\mathbf q'_{n,ll'}\cdot \mathbf r}T^n\label{eq:hopphase}\\
   T^{n-1}&=\kappa\sigma_3+\cos(2\pi n/3)\sigma_1+\sin(2\pi n/3)\sigma_2\label{eq:hopmat}
   \end{align}\label{eq:hop}
\end{subequations}
where $\mathbf q'_{n,ll'}$ are the associated tunneling wavevectors, satisfying $(\mathbf q'_{n,ll'})_x+i(\mathbf q'_{n,ll'})_y=-ik'_{\theta_{ll'}} e^{-i\frac{2\pi}{3}(n-1)}$, $k'_{\theta_{12}} = |\mathbf{K}_1 - \mathbf{K}'_2|$ and $k'_{\theta_{23}} = |\mathbf{K}'_2 - \mathbf{K}_3|$. As before, $\mathbf{K}'_2$ is defined as in Appendix \ref{appendix:geometry}. The chiral ratio, $\kappa$, suppresses intra-sublattice tunneling due to lattice relaxation and renormalization; we take $\kappa=0.7$, as many TBG studies indicate that it lies somewhere in the range 0.5--0.8~\cite{Nam_2017,Koshino_2018,carr2018relaxation,Guinea_2019,Carr_2019,Koshino_2020,Vafek_2020,ledwith2021tb,Das_2021,parker2021fieldtuned}. 
%The $\kappa = 0$, \emph{chiral} limit of this model was previously studied in Refs.~\onlinecite{foo2023extended,popov2023magic,devakul2023magic,popov2023magic2,guerci2023nature}.

% Similar to Refs.~\onlinecite{jung2014ab} and~\onlinecite{lei2021mirror}, 
We take $\mathbf q'_{n,ll'}$ to be constant in space, but allow for $\mathbf{d}$ to be spatially dependent, reflecting the super-moiré structure of the relaxed lattice~\cite{jung2014ab,lei2021mirror}. In principle, the spatially variation of $\mathbf{d}$ can be directly included in Eq.~\eqref{eq:bm} by assuming a functional form of $\mathbf{d}(\mathbf{r})$ with a commensurate supermoir\'e periodicity (this may lead to an effectively periodic supermoir\'e low-energy continuum model when the Dirac points of the moir\'e bands are dispersive, but fails when moir\'e bands are extremely flat like at the first magic angle~\cite{mao2023supermoire}). 
The main effect of incorporating the full $\mathbf{d}(\mathbf{r})$ will be to open mini-gaps in the spectrum, while greatly increasing the computational complexity of diagonalizing the Hamiltonian for large supermoir\'e periods.
These mini-gaps reflect wavefunction coherence over the supermoir\'e scale (several hundred nanometers) and are unlikely to be experimentally relevant in the presence of thermal or disorder effects (except possibly at larger angles, beyond those studied here, where the supermoir\'e period becomes smaller~\cite{zhang2021correlated})
We therefore fix $\mathbf{d}$ and treat Eq.~\eqref{eq:bm} as a local Hamiltonian that describes the moiré scale physics in a given region of the full trilayer system. 
% This Hamiltonian should capture  essential features of the electronic structure should be present at moiré scale Hamiltonian, as coherence length of the electronic wavefunctions is below the super-moiré scale. 

We use the parameters $v_F=8.8\times10^{5}\,\text{m/s}$ and tunneling energy $w=110\,\text{meV}$~\cite{bistritzer2011moire}.
% We verify the accuracy of these parameters by considering $\text{p}/\text{q} = 1$ and $\text{p}/\text{q} = -1$ TTG, which have been previously considered in other works (e.g., Refs~\onlinecite{kwan2023strong,park2021tunable}).
In Fig.~\ref{fig:DOS_vs_TwistAngle_HTG} we plot the DOS as a function of $\theta_{12}$ for $\theta_{12}/\theta_{23} = -1$ with shift $\mathbf{d} = 0$ and $\theta_{12}/\theta_{23} = 1$ with shifts $\mathbf{d} = \bm{\delta}$ (since $\mathbf{d} \rightarrow -\mathbf{d}$ under $C_{2z}\mathcal{T}$, the DOS for $\mathbf{d} = -\bm{\delta}$ is the same).
As discussed in Sec.~\ref{sec:Lattice_Relaxation}, these shifts are strongly preferred for these twist angle configurations.
We find magic angles of $1.85^\circ$ and $1.56^\circ$ respectively, in good agreement with experimental values~\cite{park2021tunable, hao2021electric, cao2021pauli, liu2022isospin,  kim2022evidence, shen2023dirac, xia2023helical}.

%CY: Should we also mention that it predicts the MQC angle? JMM: I think we will leave that for the discussion section. 
%Using the parameters, the spectrum of the local Hamiltonian in Eq. \ref{eq:bm} is only determined by the twist angles, and the moiré displacement. We assume the twist angle is constant in the system, but allow the moiré displacement to vary over the super-moiré scale. 

For the other angle ratios, we first consider the DOS as a function of twist angle. For $\theta_{12}/\theta_{23} = 1/2$, lattice relaxation results in the formation of domains with $\mathbf{d} = \pm\bm{\delta}$, and so we calculate the DOS for fixed $\mathbf{d}=\bm{\delta}$. For other $\text{p}$ and $\text{q}$, the shift varies smoothly over the super-moiré scale, such that there is no preferred shift. For these systems, we calculate the DOS averaged over all possible $\mathbf{d}$. For all cases, we estimate magic angles as the location where the DOS is most peaked near $E=0$.

%For helical TTG, we fix $\mathbf{d} = 0$ and for alternating TTG, we fix $\mathbf{d} = \bm{\delta}$. The DOS as a function of twist angle for different $\text{p}$'s and $\text{q}$'s is shown in Fig.~\ref{fig:DOS_vs_TwistAngle}. In all cases, we find magic angles where the DOS is strongly peaked. These magic angles, and others, are listed in Table~\ref{tab:MAsum}. For $|\text{p}/\text{q}| = 1/2$ and $1/3$, the DOS at the magic angle is strongly peaked in a small energy window centered at $\text{E} = 0$. For $|\text{p}/\text{q}| = 2/3$ and $3/4$, the DOS is peaked over a larger energy window, and there is significant DOS away from $\text{E} = 0$.

To understand the local physics of TTG at these magic angles, we also calculate the DOS as a function of shift at the magic angle indicated in Fig.~\ref{fig:DOS_vs_TwistAngle}. Since the supermoir\'e structure enters the local Hamiltonian through the shift, the DOS as a function of shift indicates how the local DOS varies on the supermoir\'e scale at a fixed twist angle. The DOS as a function of shift at the magic angle is shown in Fig.~\ref{fig:DOS_vs_Shift}. In Appendix~\ref{app:MABandStructure} we also plot the band structure at the magic angle for both $\mathbf{d} = 0$ and $\mathbf{d} = \bm{\delta}$. Similar to $\text{p}/\text{q} = 1$ TTG, we find gaps where the DOS vanishes above and below the central bands for $\text{p}/\text{q} = \pm 1/2$ and $\text{p}/\text{q} = -1/3$. These gaps are necessarily close when $\mathbf{d} = 0$~\cite{mora2019flatbands}, and only exist near $\mathbf{d} = \pm \bm{\delta}$.

% In general, the DOS seems to depend more heavily on shift for alternating TTG than for helical TTG, and the variance of the DOS decreases as $\text{p}$ and $\text{q}$ increase. 

\section{Discussion}\label{sec:disc}
Having presented both the lattice relaxation and electronic structure of multi-moir\'e twisted trilayer graphene, we now discuss our results and their physical consequences. 
We found that, near the first magic angle, only for $\theta_{12}/\theta_{23}\approx \pm 1$ and $1/2$ does the relaxed lattice have a periodic polycrystalline structure, consisting of large domains with a constant shift $\mathbf{d}$. 
% Our results also apply to twist angles that are slightly deviated from these commensurate fractions, as the twist angles will still relax towards the $\text{p}/\text{q}$ commensurate domains, although the size of the domains will be affected.
For all other simple rational fractions, $-1/2, \pm1/3, \pm2/3,$ and $\pm 3/4$, the supermoir\'e structure of $\mathbf{d}(\mathbf{r})$ is smoothly varying, closely resembling that of the unrelaxed structure.
These structures are therefore quasicrystalline, and do not relax towards a commensurate structure.  
This quasiperiodicity was assumed in the theoretical analysis of $(1.4^\circ,-1.9^\circ)$-TTG in Ref~\cite{uri2023superconductivity} (which is close to $-3/4$), and our full lattice relaxation result confirms this.

% For $\text{p}/\text{q} = -1$ the system has a single moiré period, however, if there is a small difference in twist angles between the layers, the system also relaxes into a polycrystalline structure. For helical TTGs, the largest domains have a shift of $\mathbf{d} = \pm \bm{\delta}$ corresponding to dominant AAB or BAA stacking, while for alternating TTG the largest domains have $\mathbf{d} = 0$ corresponding to dominant AAA stacking, in agreement with Ref.~\onlinecite{nakatsuji2023multi}. The domains are connected by narrow domain walls, where the shift quickly changes. The global physics for small $\text{p}$ and $\text{q}$ is therefore dominated by the contributions from large domains with a fixed shift and a single moiré period. 

These relaxation patterns have important implications for the electronic properties, especially for the potential realization of strongly correlated states.
For twist angles in the periodic polycrystalline regime, the ``commensurate approximation'' from Sec~\ref{sec:Lattice_Relaxation} is physical, and the system relaxes to the locally periodic structure.
For these systems, we determined the magic angle at which the DOS is maximized within the dominant domains.  
At this magic angle, polycrystalline TTG should consist of large regions with a strongly peaked DOS separated by domain walls where the DOS is lower. 
This will lead to a tendency for such materials to nucleate strongly correlated states within the domains. 
If the same correlated state can be nucleated in all domains, then proximity effects may effectively wash out the domain walls, as likely occurs in $\theta_{12}/\theta_{23}\approx -1$-TTG where large $\mathbf{d}=0$ domains form (Fig.~\ref{fig:relax11}c).
However, if the correlated states in adjacent domains have different topology, the domain walls will support gapless modes, leading to network model-like physics. 
This is believed to occur in ($1.85^\circ, 1.85^\circ$)-TTG~\cite{xia2023helical,devakul2023magic,kwan2023strong}, where symmetry-breaking phases with different Chern numbers form within the $\mathbf{d}=\bm{\delta}$ and $-\bm{\delta}$ domains. 

\begin{figure*}
    \centering
    \includegraphics[width=\textwidth]{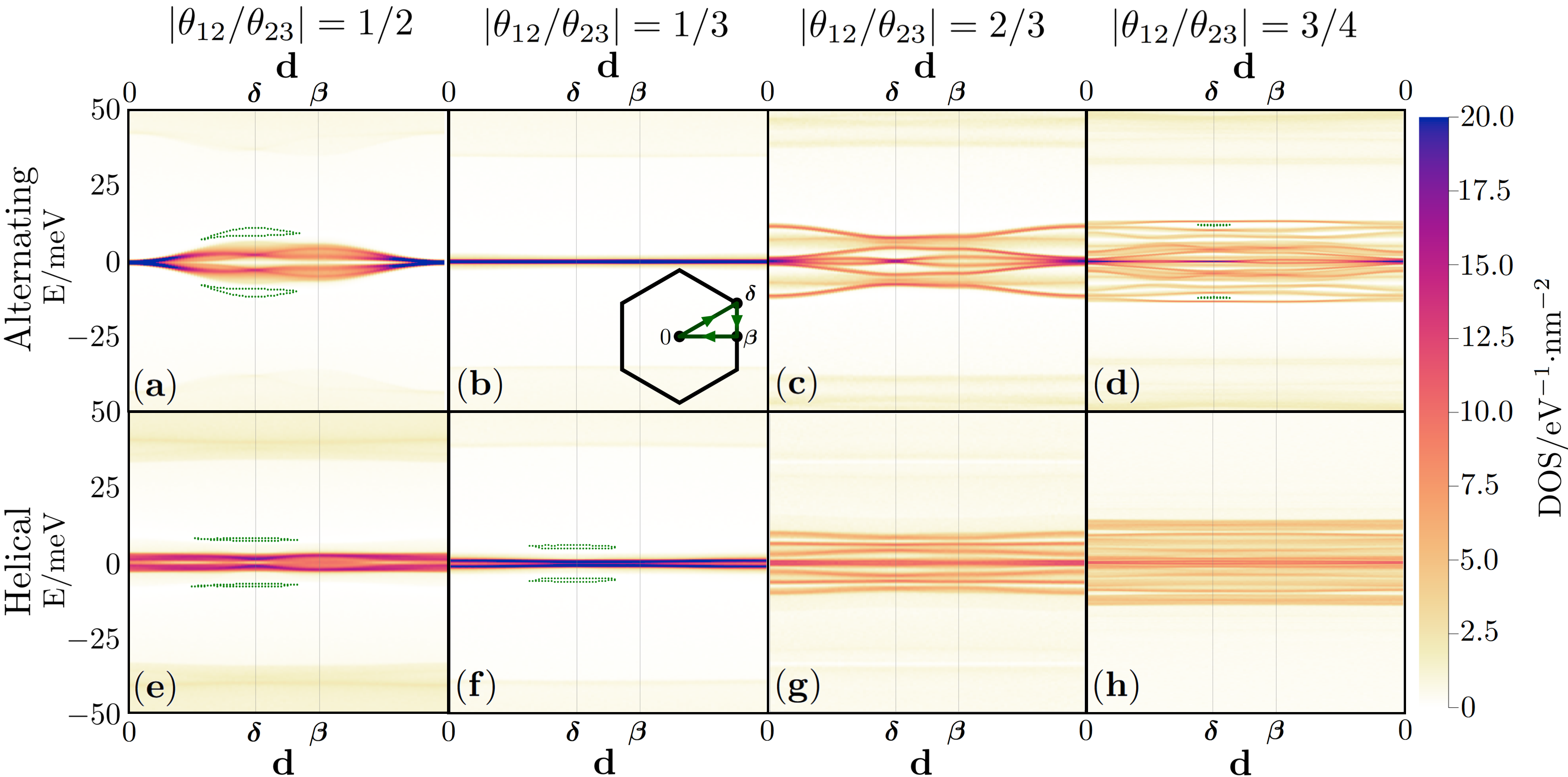}
    \caption[]{The DOS as a function of shift, $\mathbf{d}$ calculated at the magic angles shown in Table.~\ref{tab:MAsum} and Fig.~\ref{fig:DOS_vs_TwistAngle}. The path taken in real space is given in the inset to (\textbf b). 
    Dashed lines illustrate regions where the DOS is zero, indicating a full gap.}
    % The DOS is most strongly peaked at the dominant shift, $\mathbf{d} = \bm{\delta}$ for helical TTG and $\mathbf{d} = 0 $ for alternating TTG. Dotted green lines indicate regions where the DOS vanishes.} 
    \label{fig:DOS_vs_Shift}
\end{figure*}

Away from $\theta_{12}/\theta_{23}=\pm 1,1/2$, the domain walls broaden, and the system becomes less like a periodic polycrystal and more like a quasicrystal, with the relative shift smoothly changing over the supermoiré scale. 
For these cases, we determine the magic angle from the peak in the DOS averaged over $\mathbf{d}$.
Compared to $\theta_{12}/\theta_{23}=\pm 1$, Fig.~\ref{fig:DOS_vs_Shift} shows that the DOS varies less as a function of the shift in the quasi crystalline regime, and exhibits a peak for all $\mathbf{d}$.
This implies that, remarkably, the DOS is still peaked everywhere in space at the magic angle.
This should promote the creation of uniform phases, as opposed to the nucleation of local phases, although the nature of these phases may be more complicated due to quasiperiodicity.
We speculate that the manifest quasiperiodic nature of the lattice at large $\text{p}$ and $\text{q}$  may be detrimental to phases with momentum dependence, like charge density waves or non-s-wave superconductivity. 
The DOS for certain fractions also show gaps opening near $\mathbf{d} = \pm \bm{\delta}$, that close for the values of the shift (this is most pronounced in the band structures, shown in Fig.~\ref{fig:bands}, where a topological gap opens between the central flat bands and remote bands). 
If the chemical potential is in one of these gaps, the system will be gapped in the $\mathbf{d} = \pm \bm{\delta}$ regions of the relaxed structure, and a network of gapless modes will flow around them.
In general, any ``domain walls'' in the quasicrystal regime will be wide, and may therefore prefer valley-Chern or spin-Chern number changes to Chern number changes~\cite{kwan2021domain}. 

In the limit where $\theta_{12}/\theta_{23}\rightarrow 0$, 
 the quasicrystalline nature of the relaxed lattice becomes inconsequential at the first magic angle.
% the magic angle approaches the value for TBG, $\theta_{12}\approx 1.1^\circ$.
% , and the essential properties of the system \textit{only} depend on the moiré pattern of the first and second layers.
This feature can be understood by the fact that when the second and third graphene layers have a large twist angle, they are effectively electronically decoupled from each other. %, and that $\theta_{23}/\theta_{12} \rightarrow \infty$ as $\text{p}/\text{q} \rightarrow 0$.
% In this limit, the quasicrystalline nature of the relaxed lattice becomes inconsequential.
Ref.~\onlinecite{hoke2023uncovering} studied a trilayer graphene device at twist angles $\sim (1.1^\circ, -5.5^\circ)$ and showed that the third layer effectively behaved as a decoupled monolayer, while the first and second layers behaved as MATBG. 
In Table~\ref{tab:MAsum}, we find that for $\theta_{12}/\theta_{23}=1/\text{q}$, $\theta_{12}$ approaches the bilayer graphene magic angle as $|q|$ increases, with $\theta_{12} \approx 1.1^\circ$ for $|q| = 4$. 
We can understand this as the effective twist decoupling of the third layer. 
% Hence, for small $|\text{p}/\text{q}|$ it is only necessary to consider the moiré pattern formed by the first and second layers, as coupling to the third layer is highly suppressed.
Any correlated physics in this regime should match that of MATBG, with possible screening effects from the decoupled monolayer graphene. The addition of the additional monolayer graphene can also change the global symmetry of the system, but given the electronic decoupling, this explicit symmetry breaking is unlikely to be consequential.

It is worth commenting on the possible connection between the $\theta_{12}/\theta_{23}=-3/4$ TTG and the experimental observation of strongly correlated physics in the $\sim (1.4^\circ,-1.9^\circ)$ MQC~\cite{uri2023superconductivity}. 
Remarkably, our analysis suggests that $(1.4^\circ,-1.9^\circ)$ is close to precisely the magic angle of the $-3/4$ TTG system. 
The lattice relaxation calculations show a quasicystalline super-moiré structure. 
The high DOS (Fig.~\ref{fig:DOS_vs_Shift}) could also lead to flavor polarization and superconductivity observed in hole doped MQC.
However, at charge neutrality, the MQC displays quantum oscillations that are well described by dispersive Dirac cones. 
These Dirac cones are not present in the local band structure of magic angle $-3/4$ TTG calculated in this work, as shown in Fig.~\ref{fig:bands}. 
A possible resolution to this is that the Dirac velocity at the magic angle is significantly renormalized\cite{gonzalez1999marginal, mishchenko2007effect, vafek2007anomalous, park2007velocity} at charge neutrality, effectively making the flat bands Dirac-like. 
The increased DOS at the magic angle therefore either induces a correlated phase or renormalizes the system to have Dirac-like bands, depending on the filling. 
% Determining if such a scenario can occur in a multi-moir\'e system is an interesting question for future work. 
This possibility motivates the experimental study of $-3/4$ TTG at other twist angles, to determine whether strongly correlated physics only appears at our predicted magic angle of $(1.41^\circ,-1.88^\circ)$, or whether it appears more generically.

In this work we have focused on the properties of multi-moir\'e TTG near twist angles with simple fractions $\theta_{12}/\theta_{23}=\text{p}/\text{q}$ in order to probe the tendency to relax into exactly commensurate structures. 
However, our results apply more generally. 
% Here, the exact values of $\text{p}$ and $\text{p}$ only determine the size of the moiré Brillouin zone. Having a periodic moiré Brillouin zone is useful for calculations of the DOS, but the DOS itself does not depend directly on the existence of a moiré Brillouin zone. 
% Importantly, the DOS is a smooth function of the twist angles, \textit{not} of $\text{p}$ and $\text{q}$.  
Ultimately, the DOS is a physical quantity which should depend smoothly on the twist angles.
Angles that lie on the magic line in Fig.~\ref{fig:TTG_Datapoints} will have a high DOS, even if they are not exactly rational. 
Our conclusions of the relaxed lattice structure will also depend smoothly on the twist angles. 
We therefore conclude that all along the magic line, multi-moir\'e TTG will consist of regions where the relaxed lattice structure is a quasicrystal, with small regions near $(1.56^\circ, -1.56^\circ)$, $(1.85^\circ, 1.85^\circ)$, and $(1.31^\circ, 2.61^\circ)$ where the relaxed lattice structure is a periodic polycrystal, and twist decoupling in the large $\theta_{23}$ limit. 
Notably, the periodic polycrystals are stable to small changes in the twist angle --- the main effect of a small twist detuning will merely affect size of the domains~\cite{nakatsuji2023multi,devakul2023magic,xia2023helical}. 
Thus, the behaviors identified in this work are therefore not tied to specific rational twist angles, but are part of a continuum of TTG systems along a magic line. 

Our results have important implications for future experiments.  
Our results for the relaxed structure, in particular the formation or non-formation of commensurate domains, can be readily probed by imaging techniques such as atomic force microscopy or scanning tunneling microscopy (STM).
STM can also directly probe the local DOS as a function of position; for quasicrystalline structures, $\mathbf{d}$ varies smoothly in space and our DOS results shown in Fig.~\ref{fig:DOS_vs_Shift} directly reflects the supermoir\'e scale spatial dependence of the local DOS.
For twist angles where the supermoir\'e periods are very large, on the order of hundreds of nanometers, techniques such as scanning single-electron transistor or microwave impedance microscopy can detect the supermoir\'e scale spatial variations in the electronic properties.
In transport, we predict that correlated behaviors are likely to appear along a magic line where the DOS is maximized.  
This physics may manifest in a variety of different ways: correlated insulators, orbital ferromagnetism, anomalous Hall effects, or superconductivity, to name some possibilities.
The broad variety of physics already observed in experiments~\cite{park2022robust, zhang2021ascendance,park2021tunable, hao2021electric, cao2021pauli, liu2022isospin,  kim2022evidence,uri2023superconductivity,hoke2023uncovering,xia2023helical} is evidence of the wide range of possibilities in this platform.
The precise nature of the correlated physics that will appear in each system along the magic line is an interesting open question for future theoretical and experimental studies.

In summary, we have investigated the lattice and electronic properties of multi-moiré TTG near the first magic angles. 
The combined results of our analysis of the lattice relaxation and electronic properties of multi-moir\'e TTG reveals a rich phase diagram of polycrystalline and quasicrystalline structures along a magic line of twist angles where the DOS is sharply peaked, summarized in Fig.~\ref{fig:TTG_Datapoints}.
At these twist angles, we predict the realization of strongly correlated physics.
Many interesting strongly correlated phenomena have already been observed in multi-moir\'e TTG systems (Table~\ref{tab:MAsum}); however, the full phase space is only just beginning to be explored.
Our study sets the stage for future theoretical and experimental exploration of the full phase space of multi-moir\'e TTG.

% Our analysis reveals a rich phase diagram of periodic polycrystalline and quasicrystalline lattice structures, reveals several qualitatively different regimes, summarized in Fig.~\ref{fig:TTG_Datapoints}
% We identify angles at which the DOS is sharply peaked, near which we predict strongly correlated physics to appear experimentally.
% The rational twist angles we find her make up a magic line of twist angles, made up of regions where the super-moire structure is either a periodic-polycrystal or a quasicrystal.
% Far along the magic line, where the relaxed lattice structure in nominally a one quasicrystal, one of the twist angles becomes large enough to decouple one of the layers from the other two.
% In this limit, the system only depends on the single moiré period set by the first and second layers, and resembles MATBG.
% The line of magic angles found in this work are therefore ideal places to search for a wide array of different correlated behavior in experimental studies.
% More theoretical work is also needed to determine what types of correlated phases are expected in different regimes, and how this relates to the polycrystal or quasicrystal structure of the relaxed lattice. 

\begin{acknowledgments}
We thank Aviram Uri, Liqiao Xia, Aaron Sharpe, and Sergio de la Barrera for helpful discussions leading to this project and for valuable comments on the manuscript. 
Z.Z. is supported by a Stanford Science Fellowship. 
This work is supported by a startup fund at Stanford University.
\end{acknowledgments}

\bibliography{NCTTG.bib}
\bibliographystyle{apsrev4-1}

\onecolumngrid
\newpage{}
\appendix
\section{Local Commensuration}\label{appendix:geometry}
We denote the lattice vectors for individual layers as $\mathbf A_{l}$, the moir\'e lattice as $\mathbf A_{ll'}$ and the moire-of-moire lattice vectors as $\mathbf A$. The corresponding reciprocal lattice vectors are $\mathbf G_{l}$, $\mathbf G_{ll'}$ and $\mathbf G$. Primed vectors denote the distorted, commensurate vectors. These will denote the matrices $\mathbf G_{l} = [\mathbf g_{1,l}, \mathbf g_{2,l}] = [\mathbf g_{1,l}, \mathbf R(2\pi/3)\mathbf g_{1,l}] $, and so forth, where $\mathbf R(\theta)$ is a rotation matrix.

If we wish to enforce global commensuration
\begin{equation}
\text{q}\mathbf G_{12}' = \text{p}\mathbf G_{23}'
\end{equation}
we can fix two of the layers and distort the third. We choose to distort the second layer as it is the most symmetric distortion. We do so by using the fact that the moir\'e reciprocal vectors can be written in terms of the layer lattice vectors
\begin{equation}
 \mathbf G_{ll'}=\mathbf G_{l'}-\mathbf G_{l}
\end{equation}

Since $\mathbf g_{2,ll'}, \mathbf g_{2,l}$ can be written as an appropriate rotation of $\mathbf g_{1,ll'}, \mathbf g_{1,l}$, we can distort the second layer in a way that preserves its hexagonal symmetry. This particular distortion is given by the condition
\begin{equation}
    \mathbf G_{2}' = \frac{\text{q}}{\text{p}+\text{q}}\mathbf G_{1}+\frac{\text{p}}{\text{p}+\text{q}}\mathbf G_{3}
\end{equation}
which gives us additionally our hopping wavevectors and our supermoir\'e reciprocal vectors
\begin{subequations}
    \begin{align}
        \mathbf G_{12}' &= \frac{\text{p}}{\text{p}+\text{q}}(\mathbf G_{3}-\mathbf G_{1})\label{eq:dist12}\\
        \mathbf G_{23}' &= \frac{\text{q}}{\text{p}+\text{q}}(\mathbf G_{3}-\mathbf G_{1})\label{eq:dist23}\\
        \mathbf G' &=\frac{1}{p+q}(\mathbf G_{3}-\mathbf G_{1})\label{eq:distmm}
    \end{align}\label{eq:dist}
\end{subequations}
We visualize this distortion as shifting the $\bm K_2$ point onto the line connecting $\bm K_1$ to $\bm K_3$, as seen in Fig.~\ref{fig:distorion}.
Naturally, Eqs.~\ref{eq:dist} diverge for (p,q)=(1,-1), but since the moir\'e lattices are already exactly commensurate, we don't need to use this procedure. These reciprocal vectors encode everything about the geometry of the system, and all computations are done in terms of these reciprocal vectors.
From these distorted reciprocal vectors, the distorted real space lattice vectors can be calculated
\begin{subequations}
\begin{align}
    \mathbf A_l'= 2\pi[ (\mathbf G_l')^\intercal]^{-1}\\
    \mathbf A_{ll'}'= 2\pi[ (\mathbf G_{ll'}')^\intercal]^{-1}\\
    \mathbf A'= 2\pi[ (\mathbf G')^\intercal]^{-1}
    \end{align}
\end{subequations}
Since it is shared in both moir\'e BZs, we fix $\mathbf K_2'$ at the centre of the supermoir\'e BZ, $\gamma$, and position $\mathbf K_1 = \mathbf K_2'+\mathbf q'_{1,12}$ and $\mathbf K_3=\mathbf K_2' -\mathbf q_{1,23}'$ accordingly. Since we enforce exact commensuration, these will fold onto one of the high symmetry points $\gamma$, $\kappa$, $\kappa'$, depending on the exact structure. Since we neglect Pauli rotation of the Dirac cones, we ensure that the reciprocal vector \(\mathbf G'\) is exactly aligned with the coordinate axis, which maintains exact particle-hole symmetry of the DOS.

\section{Magic Angle Band Structures}\label{app:MABandStructure}
\begin{figure*}
    \centering
    \includegraphics[width=\textwidth]{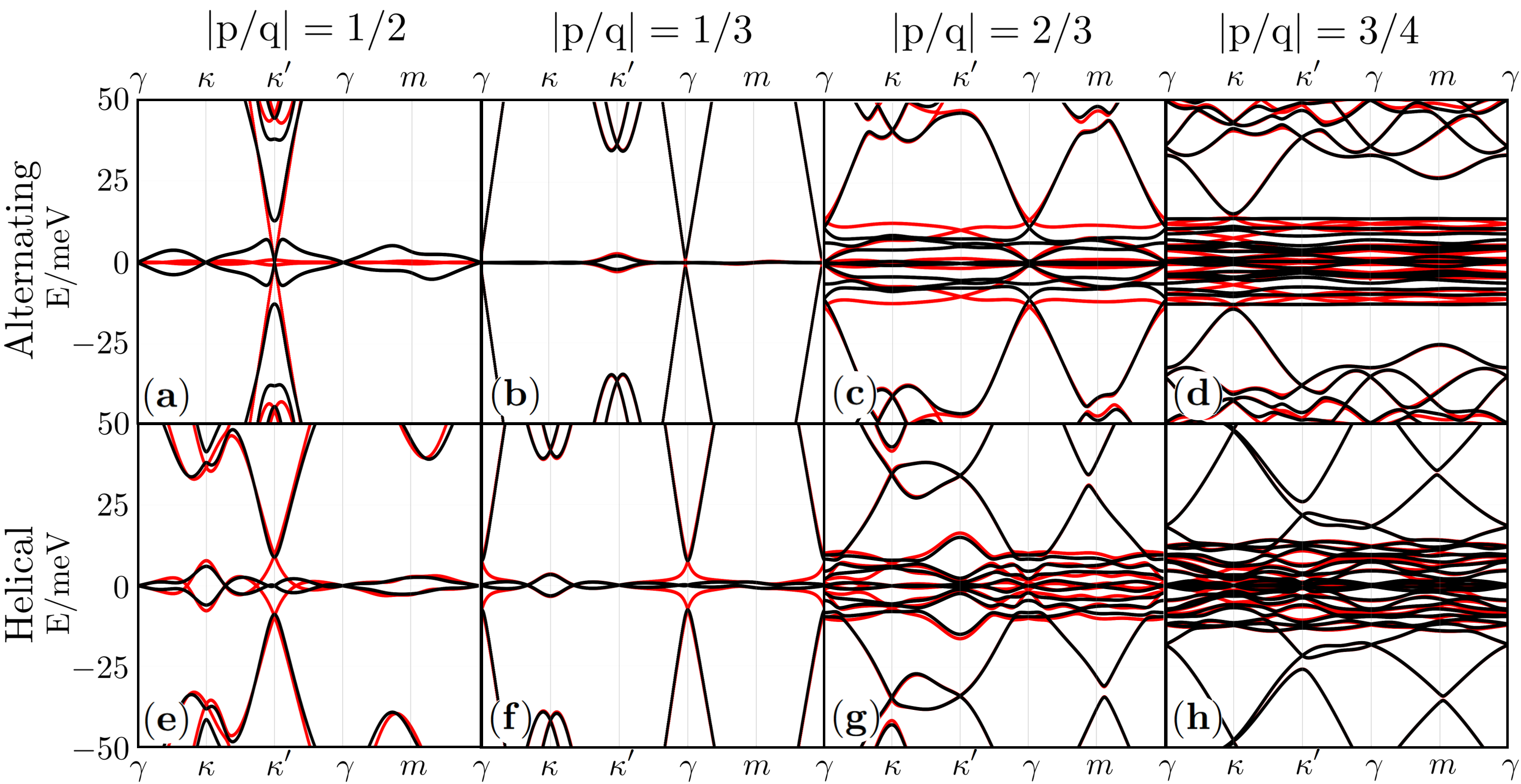}
    \caption{Band structures near magic angles for selected configurations.
    The black bands represent the shift $\mathbf d= \bm \delta$, while the red bands represent the shift $\mathbf d = 0$
    Note that in (\textbf{a},\textbf{e}, \textbf{f}) that the central bands are gapped at $\mathbf d = \bm \delta$, as illustrated in Fig.~\ref{fig:DOS_vs_Shift}.
    } \label{fig:bands}
\end{figure*}

We begin with Eq.~\eqref{eq:bm}, dropping the Pauli rotation ($\bm\sigma_{\theta}\to\bm\sigma$).
%Implicitly, the Hamiltonian has been written in terms of shifted momenta so that the kinetic terms on the diagonal are measured with respect to the layer's $K_l$ point. We make this more explicit (and simultaneously drop Pauli rotation)
%\begin{equation}
%    H_{\bm K}=\begin{pmatrix}
%        -iv_F \bm\sigma\cdot(\bm\nabla-\bm K_1) & T_{12}(\mathbf r - \mathbf d_t) & 0\\
%        T^\dagger_{12}(\mathbf r - \mathbf d_t) & -iv_F\bm\sigma\cdot(\bm\nabla-\bm K_2) & T_{23}(\mathbf r - \mathbf d_b)\\
%        0 & T^\dagger_{23}(\mathbf r - \mathbf d_b) & -iv_F\bm\sigma\cdot(\bm\nabla-\bm K_3)
%\end{pmatrix}\label{eq:kbm}
%\end{equation}
We apply a unitary transformation to shift the tunneling wavevectors by $-\mathbf q_{0,ll'}'$, so that we may transform 
\begin{equation}
    \mathbf q_{0,ll'}', \mathbf q_{1,ll'}', \mathbf q_{2,ll'}'\longrightarrow 0, -\mathbf g_{1,ll'}', \mathbf g_{2,ll'}'
\end{equation}
explicitly, this unitary for shifting $\text{q}'_{n,ll'}$ is given
\begin{subequations}
    \begin{align}
U_{12}&=\begin{pmatrix}
    e^{-i\mathbf q'_{0,12}\cdot\mathbf r} & 0 & 0\\
    0 & e^{i\mathbf q'_{0,12}\cdot\mathbf r} & 0\\
    0 & 0 & 1
\end{pmatrix}\\
U_{23}&=\begin{pmatrix}
    1 & 0 & 0\\
    0 & e^{-i\mathbf q'_{0,23}\cdot\mathbf r} & 0 \\
    0 & 0 & e^{i\mathbf q'_{0,23}\cdot\mathbf r} \\
\end{pmatrix}
    \end{align}
\end{subequations}
so that
\begin{equation}
    U_{12}U_{23}H_{\bm K}U_{23}^\dagger U_{12}^\dagger = \begin{pmatrix}
        -iv_F \bm\sigma\cdot(\bm\nabla-\bm K_1) & T_{12}(\mathbf r - \mathbf d_t)e^{-i\mathbf q'_{0,12}} & 0\\
        T^\dagger_{12}(\mathbf r - \mathbf d_t)e^{i\mathbf q'_{0,12}} & -iv_F\bm\sigma\cdot(\bm\nabla-\bm K_2) & T_{23}(\mathbf r - \mathbf d_b)e^{-i\mathbf q'_{0,23}}\\
        0 & T^\dagger_{23}(\mathbf r - \mathbf d_b)e^{i\mathbf q'_{0,23}} & -iv_F\bm\sigma\cdot(\bm\nabla-\bm K_3)
    \end{pmatrix}\label{eq:bmu}
\end{equation}
The unitaries also affect the diagonal kinetic terms as a momentum shift. Then, we can rewrite
\begin{equation}
    \tilde T_{ll'}(\mathbf r) =w\sum_{n=0}^2 e^{i\mathbf g'_{n,ll'}\cdot \mathbf r}T^n 
\end{equation}
where we define $\mathbf g'_{0,ll'}=0$ for convenience. 
%The unitary transform~\ref{eq:bmu} simplifies the Bloch periodicity of wavefunctions on the $l$th layer. 
% For~\ref{eq:bm}, the periodicity depends on the layer\cite{devakul2023magic}
% \begin{equation}
%     \psi_{\mathbf k,l}(\mathbf r+\mathbf a)=e^{i(\mathbf k-\bm K_l)\cdot\mathbf a}\psi_{\mathbf k,l}(\mathbf r)
% \end{equation}
%whereas 
The wavefunctions of the transformed Hamiltonian satisfies the Bloch periodicity
\begin{equation}
    \psi'_{\mathbf k}(\mathbf r+\mathbf a)=e^{i\mathbf k \cdot \mathbf a}\psi'_{\mathbf k}(\mathbf r)
\end{equation}
Since we have approximated our moir\'e lattices to be exactly commensurate, this allows us to define a unified BZ that is labelled by the momenta of all layers simultaneously.

Thus, we choose as a basis planewave states $\lvert \mathbf k, l, \tau\rangle$ labeled by crystal momentum $\mathbf k$, layer $l$, and sublattice $\tau$. These satisfy
\begin{equation}
    \langle \mathbf r \vert \mathbf k, l, \tau\rangle = e^{i\mathbf k \cdot \mathbf r}\lvert l,\tau\rangle
\end{equation}
so
\begin{equation}
    \langle\mathbf r\rvert e^{i\mathbf q \cdot \mathbf r}\lvert \mathbf k, l, \tau\rangle = e^{i(\mathbf k+\mathbf q) \cdot \mathbf r}\lvert l,\tau\rangle = \langle \mathbf r \vert \mathbf k + \mathbf q, l,\tau\rangle
\end{equation}
Thus, we see that multiplying by a planewave serves to couple different \(\mathbf k\) modes; this is the effect of the operator $e^{i\mathbf g_{n,ll'}'\cdot\mathbf r}$ in the expression for \(\tilde T_{ll'}\). 

In this basis, the Hamiltonian has layer block-diagonal entries
\begin{equation}
    \langle \mathbf k,l,\tau \rvert H_{\bm K}\lvert \mathbf k,l,\tau' \rangle = v_F\left[\sigma\cdot(\mathbf k - \bm K_l)\right]_{\tau\tau'}
\end{equation}
and off-diagonal blocks
\begin{equation}
    \langle\mathbf k, l,\tau \rvert H_{\mathbf K}\lvert\mathbf k',l',\tau'\rangle=\sum_{n=0,1,2}\sum_{\xi=\pm1} \delta_{\mathbf k', \mathbf k+\mathbf g_{n,ll'}'}we^{i\xi\mathbf g_{n,ll'}'\cdot\mathbf d_{ll'}}T^n_{\tau\tau'}\delta_{l',l+\xi}
\end{equation}
where we define for convenience $g_{n,ll'}' = -g_{n,l'l}'$ and $d_{ll'}=-d_{ll'}$, and we have rewritten $d_{12}\equiv d_t$ and $d_{23}=d_b$. 

Using the moir\'e reciprocal vectors $\mathbf g_{n,12}',\,\mathbf g_{n,23}'$ as the new hopping vectors, a $k$-space lattice representing the Hamiltonian and its couplings is constructed out to a cutoff $E/\hbar v_F$ which is set to minimize finite size effects while keeping memory usage reasonable. As the energy cutoff $E$ increases, we consider momenta with increasingly higher momenta which couple more weakly to the momenta within the supermoir\'e BZ. To calculate the band structures, a consistent $E=3000\,\text{meV}$ was used; for the DOS plots, we used a cutoff of $E=3500\,\text{meV}$ for $\lvert \text{p}/\text{q} \rvert = 1$, $E=3300\,\text{meV}$ for $\lvert \text{p}/\text{q} \rvert = 1/2,\,1/3,\,2/3$, and $E=3000\,\text{meV}$ for $\lvert \text{p}/\text{q} \rvert = 3/4$, 
This Hamiltonian is numerically diagonalized within the supermoir\'e BZ using the \texttt{ArnoldiMethod.jl} package and using shift-and-invert transformations to target the eigenvalues closest to zero. To highlight the effect of moir\'e shift on the band structure (or indeed, lack thereof), we plot the band structure at each magic angle for the two high-symmetry shifts ($\mathbf d = 0,\bm\delta$) in Fig.~\ref{fig:bands}. 

\section{Density of States}
The density of states is calculated by randomly sampling the Brillouin zone and finding the energy eigenvalues. These energies are converted to the density of states by using a Gaussian as an approximation for the delta function, setting the standard deviation to be the same as the width of the energy bins plotted. The density of states is normalized by the area of the supermoir\'e unit cell. For example, in (1,q) structures, filling the two flat bands represents eight electrons in the moir\'e  unit cell of the smaller angle, or in $\text{q}^2$ moir\'e unit cells of the larger angle. The magic angle is obtained by plotting the density of states against the smaller twist angle, and extracting the angle where the bands are most constricted and show the greatest density of states near \(E=0\). For structures that show strong relaxation, we compute the density of states by fixing the shift to be the dominant shift of the relaxed domains. For structures with weaker relaxation, we randomly sample the shifts in addition to the momenta to build up an averaged density of states.

Similarly, we plot the density of states as a function of moir\'e shifts to illuminate the effects of the spatially varying shift on the local density of states. We plot the density of states as slices along a real space path. Since approximating the delta functions as a Gaussian blurs the boundaries for gap openings, we set a small cutoff $\sim10^{-4}-10^{-8}\,\text{eV}^{-1}\cdot\text{nm}^{-2}$ below which we consider to be a gap. As seen in the black band structures in Fig.~\ref{fig:bands}(\textbf{a},\textbf{e},\textbf{f}), these regions do indicate true gap openings.

\section{Configuration space continuum model for relaxation patterns}~\label{sec:relax}
To calculate the atomic relaxation pattern, we employ a continuum relaxation model in local configuration space~\cite{zhu2020modeling}.
In twisted trilayer graphene with two independent twist angles, there does not exist a largest length scale and the system is incommensurate~\cite{zhu2020modeling}.
Therefore, instead of formulating the problem in real space, we adopt configuration space, which describes the local environment of every position in layer $L_\ell$ and bypasses a periodic approximation~\cite{cazeaux2020energy}. 
Every position in real space $\vec{r}$ in $L_i$ can be uniquely parametrized by three shift vectors $\vec{b}^{i\rightarrow j}$ for $j = 1, 2, 3$ that describes the relative position between a given point $\vec{r}$ with respect to all three layers. Note that $\vec{b}^{i \rightarrow j} = \vec{0}$ if $i = j$ since the separation between a position with itself is 0, which leads to a four-dimensional configuration space. 

% mapping between configuration space and real space
For a given real space position $\vec{r}$, the following linear transformation relates $\vec{r}$ and $\vec{b}^{i\rightarrow j}$ in layer $i$ with respect to layer $j$, 
and the following linear transformation maps the relaxation from the local configuration to the real space position $\bm r$:
\begin{equation}
\bm b^{i\rightarrow j} (\vec{r}) = (E_j^{-1} E_i - \mathbb{1}) \bm r, \label{eqn:mapping}
\end{equation}
where $E_i$ and $E_j$ are the unit cell vectors of layers $i$ and $j$ respectively, rotated by $\theta_{ij}$. In the trilayer system, there is no simple linear transformation between real and configuration space. The relation between the displacement field defined in real space, $\vec{U}^{(i)} (\vec{r})$, and in configuration space, $\vec{u}^{(i)} (\vec{b}) $, can be found by evaluating $\vec{u}^{(j)} (\vec{b}) $ at the corresponding $\vec{b}^{i\rightarrow j} (\vec{r})$ and $\vec{b}^{i\rightarrow k} (\vec{r}) $ with Eq.~\eqref{eqn:mapping} to obtain 
\begin{equation}
    \vec{U}^{(i)} (\vec{r}) = \vec{u}^{(i)}(\vec{b}^{i\rightarrow j} (\vec{r}), \vec{b}^{i\rightarrow k} (\vec{r})), 
\end{equation}
where $j, k \neq i$ and $j < k$.

The relaxed energy has two contributions, intralayer and interlayer energies:
\begin{align}
	&E^\rr{tot} (\bu^{(1)}, \bu^{(2)}, \bu^{(3)}) = E^\rr{intra} (\bu^{(1)}, \bu^{(2)}, \bu^{(3)}) + E^\rr{inter} (\bu^{(1)}, \bu^{(2)}, \bu^{(3)}), \label{eq:total_E}
\end{align}
where $\bu^{(\ell)}$ is the relaxation displacement vector in layer $\ell$. 
To obtain the relaxation pattern, we minimize the total energy with respect to the relaxation displacement vector. 

We model the intralayer coupling based on linear elasticity theory:
\begin{align}
E^\mathrm{intra} (\bm u^{(1)}, \bm u^{(2)}, \bm u^{(3)}) 
&= \sum_{\ell=1}^3 \int \frac{1}{2} \Big[G (\partial_x u^{(\ell)}_x + \partial_y u^{(\ell)}_y)^2  \nonumber \\
&\ \ \  + K ( (\partial_x u^{(\ell)}_x - \partial_y u^{(\ell)}_y)^2 + (\partial _x u^{(\ell)}_y + \partial_y u^{(\ell)}_x)^2) \Big] d \vec{b},\label{eqn:intra}
\end{align}
where $G$ and $K$ are shear and bulk moduli of monolayer graphene, which we take to be $G = 47352 \, \mathrm{meV/unit \ cell}$, $K = 69518 \, \mathrm{meV/unit \ cell}$~\cite{carr2018relaxation,zhu2020modeling}.  
Note that the gradient in Eq.~\eqref{eqn:intra} is with respect to the real space position $\vec{r}$. 

The interlayer energy accounts for the energy cost of the layer misfit, which is described by the generalized stacking fault energy (GSFE) ~\cite{kaxiras1993free,zhou2015vdw}, obtained using first principles Density Functional Theory (DFT) with the Vienna Ab initio Simulation Package (VASP)~\cite{Kresse1993, Kresse1996a, Kresse1996b}. GSFE is the ground state energy as a function of the local stacking with respect to the lowest energy stacking between a bilayer. For bilayer graphene, GSFE is maximized at the AA stacking and minimized at the AB stacking. Letting $\vec{b} = (b_x, b_y)$ be the relative stacking between two layers, we define the following vector $\bm v= (v, w) \in [0, 2 \pi ]^2$:
\begin{equation}
	\begin{pmatrix}
		v \\ w
	\end{pmatrix}
		= \frac{2 \pi}{a_0} \mqty[\sqrt{3}/2 & -1/2 \\ \sqrt{3}/2 & 1/2]
	\begin{pmatrix}
		b_x \\ b_y
	\end{pmatrix},
\end{equation}
where $a_0 = 2.4595 \text{\AA}$ is the graphene lattice constant. 
We parametrize the GSFE as follows, 
\begin{align}
V^\mathrm{GSFE}_{j\pm} = c_0 + & c_1(\cos v + \cos w + \cos (v + w) ) \nonumber \\
+ & c_2 (\cos (v + 2w) + \cos(v-w) + \cos(2 v + w)) \nonumber \\
+ & c_3 (\cos(2 v ) + \cos (2 w) + \cos(2 v + 2 w)),\label{eqn:vgsfe}
\end{align}
where we take $c_0 = 6.832\, \mathrm{meV/cell}$, $c_1 = 4.064 \, \mathrm{meV/cell}$, $c_2 = -0.374 \, \mathrm{meV/cell}$, $c_3 = -0.0095\, \mathrm{meV/cell}$~\cite{zhu2020modeling,carr2018relaxation}.
The van der Waals force is implemented through the vdW-DFT method using the SCAN+rVV10 functional~\cite{Peng2016}. In terms of $V^\mathrm{GSFE}_{\ell\pm}$, the total interlayer energy can be expressed as follows:
\begin{align}
	E^\mathrm{inter} &= \frac{1}{2}\int \misfitr{1+}{1}{2}\,\mathrm{d} \vec{b} +\frac{1}{2}\int \left[ \misfitr{2-}{2}{1} + \misfitr{2+}{2}{3} \right]\,\mathrm{d}\vec{b} \nonumber \\
&+\frac{1}{2}\int \misfitr{3-}{3}{2} \,\mathrm{d}\vec{b}, \nonumber
\end{align}
where $\vec{B}^{i\rightarrow j} = \vec{b}^{i\rightarrow j} + \bu^{(j)} - \bu^{(i)}$ is the relaxation modified local shift vector.
Note that we neglect the interlayer coupling between layers 1 and 3. 
The total energy is obtained by summing over uniformly sampled configuration space. The discretization of the four-dimensional configuration space depends on the twist angle. 
In general, the larger the moir\'e of moir\'e periodicity is, the more dense the sampling needs to be in order to resolve the large scale patterns. 
For the $(1.60^\circ, 1.65^\circ)$ case, we choose the discretization to be $54 \times 54 \times 54 \times 54$;
% for the $(0.6^\circ, 1.2^\circ)$ case, we choose $x \times x \times x \times x$, 
and for rest of the cases, 
we choose $36 \times 36 \times 36 \times 36$. 

\section{AA stacking regions of the fully relaxed structure}
\begin{figure}
    \centering
    \includegraphics[width=0.5\linewidth]{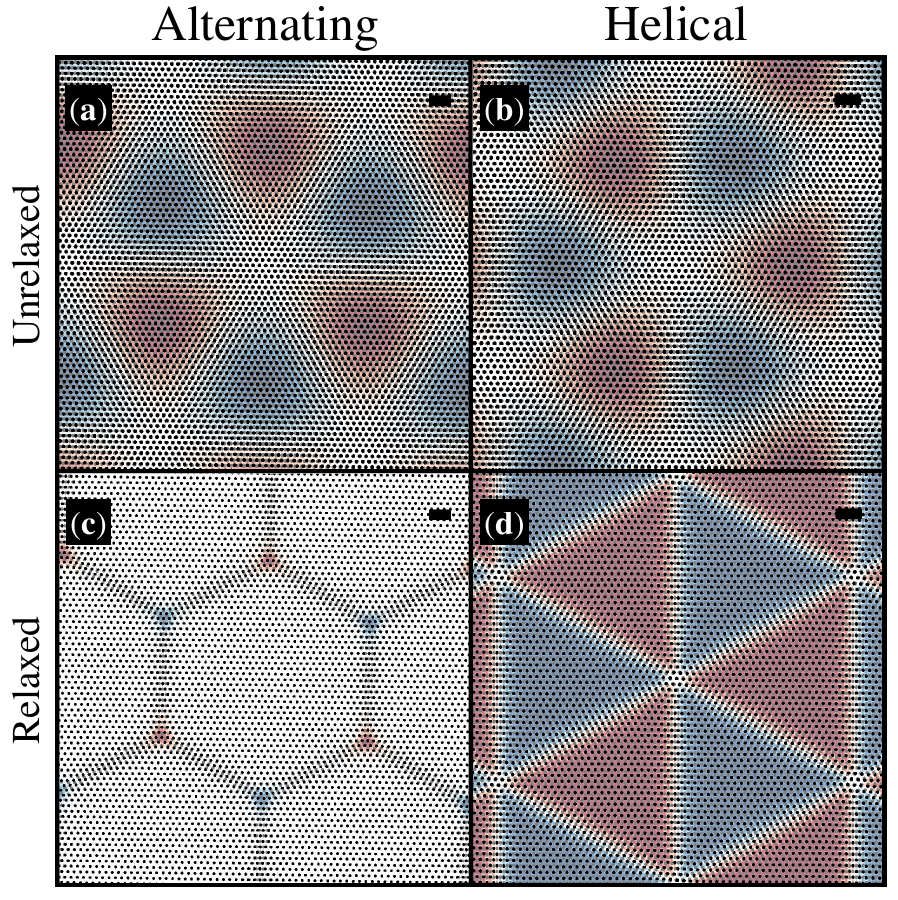}
    \caption{ 
    The relaxed structure for the first magic angles are shown,
    for the case of \textbf{(a)} almost mirror symmetric  $(1.60^\circ,-1.65^\circ)$-TTG and \textbf{(b)} $(1.85^\circ,1.85^\circ)$-TTG.
    The gray and black dots represent the AA stacking regions of layers 1,2, and layers 2,3, respectively. 
    The function $W(\mathbf{d}(\mathbf{r}))$ from the main text is plotted in the background.  The scalebar represents 30nm.
    }
    \label{fig:relax11_dots}
\end{figure}

\begin{figure*}
    \centering
    \includegraphics[width=1\linewidth]{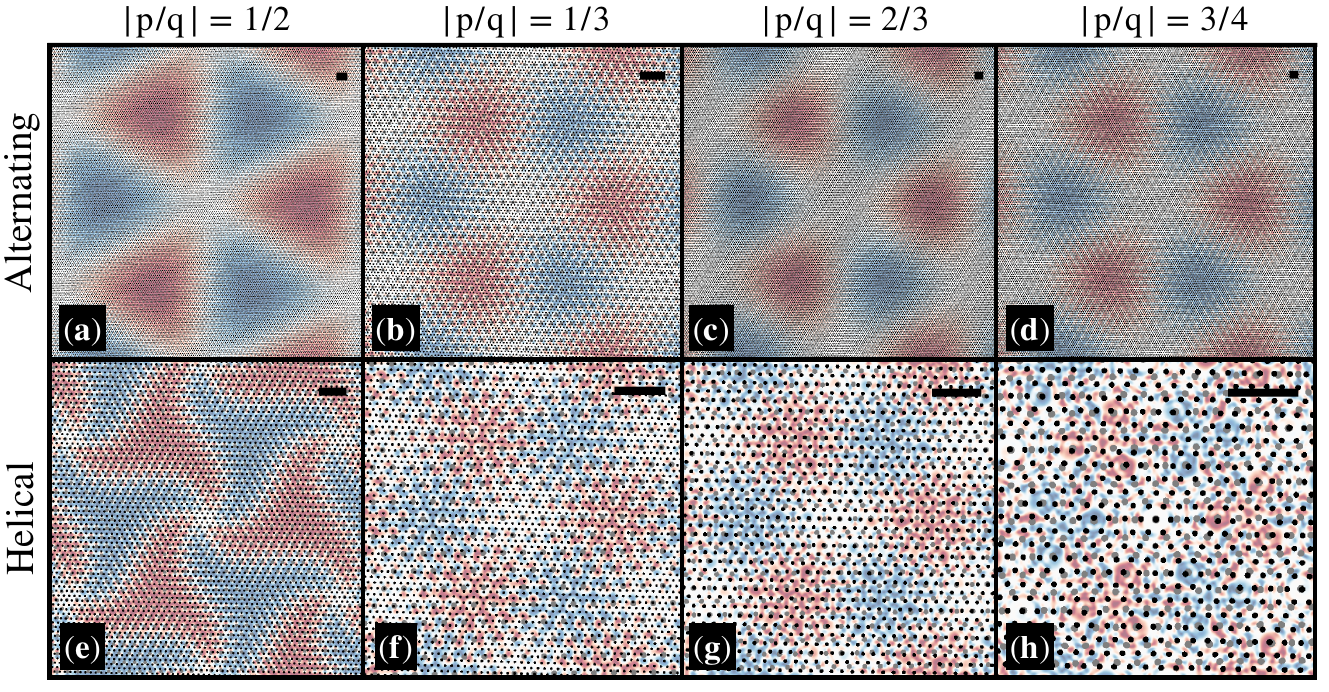}
    \caption{ 
    The relaxed structure for the first magic angles are shown, for several rational twist angle configurations.  
    The gray and black dots represent the AA stacking regions of layers 1,2, and layers 2,3, respectively. 
    The function $W(\mathbf{d}(\mathbf{r}))$ from the main text is plotted in the background. The twist angles are \textbf{(a)} $(1.25^\circ, 2.50^\circ)$,  \textbf{(b)} $(1.20^\circ, 3.60^\circ)$, \textbf{(c)} $(1.50^\circ, 2.25^\circ)$, \textbf{(d)} $(1.60^\circ, 2.13^\circ)$, \textbf{(e)} $(1.30^\circ, -2.60^\circ)$, \textbf{(f)} $(1.20^\circ, -3.60^\circ)$, \textbf{(g)} $(1.40^\circ, -2.10^\circ)$, and \textbf{(h)} $(1.50^\circ, -2.00^\circ)$. The scalebar represents 30nm.
    }
    \label{fig:relax_general_dots}
\end{figure*}

In the main text, we showed the shift vector $\mathbf{d}(\mathbf{r})$ of the relaxed structure.
In Figures~\ref{fig:relax11_dots} and \ref{fig:relax_general_dots}, we further show the AA stacking regions of the fully relaxed structures, providing more insight into the precise relaxation pattern in each of these structures.

\section{Shift vector derivation}\label{sec:dvec}
In this section, we derive the expression for the shift vector $\mathbf{d}(\mathbf{s}_1,\mathbf{s}_2,\mathbf{s}_3)$, as a function of the local atomic shifts $\mathbf{s}_l$, stated in the main text.  
For concreteness, let us denote $\mathbf{A}_0$ the matrix 
\begin{equation}
\mathbf{A}_0=a_0\begin{pmatrix} 1 & \frac{1}{2}\\ 0 & \frac{\sqrt{3}}{2}\end{pmatrix}
\end{equation}
with columns given by the lattice vectors of (untwisted) graphene, and $a_0=2.46$\AA.

Let us consider a structure in which three layers are rigidly twisted about the origin by angles $\theta_l$, and shifted by $\mathbf{s}_l$.
The lattice vectors of layer $l$ is $\mathbf{A}_l=\mathbf{R}(\theta_l)\mathbf{A}_0$, where $\mathbf{R}(\theta)$ is the rotation matrix. 
Note that $\mathbf{s}_l$ is only well defined modulo $\mathbf{A}_l$.  

Now, suppose that we slightly deform the lattice $\mathbf{A}_2\rightarrow\mathbf{A}_2^\prime$ about the origin.
This deformation sends $\mathbf{s}_2\rightarrow\mathbf{s}_2^\prime=\mathbf{A}_2^\prime\mathbf{A}_2^{-1}\mathbf{s}_2$.
Note that $\mathbf{s}_2^\prime$ is well defined modulo $\mathbf{A}_2^\prime$.

After this deformation, the moir\'e unit cells of layers 1,2,  and 2,3, are exactly commensurate.  
Let us denote the moir\'e unit vectors
\begin{equation}
\begin{split}
\mathbf{A}_{12}^\prime=(\mathbf{A}_1^{-1}-\mathbf{A}_2^{\prime-1})^{-1}\\
\mathbf{A}_{23}^\prime=(\mathbf{A}_2^{\prime-1}-\mathbf{A}_3^{-1})^{-1}
\end{split}
\end{equation}
which are commensurate with the enlarged unit cell $\mathbf{A}^\prime\equiv p\mathbf{A}_{12}^\prime=q\mathbf{A}_{23}^\prime$.

Let us determine $\mathbf{d}_{12}$, the position of an AA site of layers $1$ and $2$ within the commensurate structure.
We define an AA site to be the location where the local shift of the two layers, within their respective unit cells, is identical.  
At a position $\mathbf{r}$ away from the origin, the local shift appears modified by $-\mathbf{r}$.  
Thus, $\mathbf{d}_{12}$ is determined by the equation
\begin{equation}
\mathbf{A}_{1}^{-1}(\mathbf{s}_1-\mathbf{d}_{12}) = \mathbf{A}_2^{\prime-1}(\mathbf{s}_2^\prime-\mathbf{d}_{12})
\end{equation}
implying
% \begin{equation}
% (\mathbf{A}_{2}^{\prime-1}-\mathbf{A}_1^{-1})\mathbf{d}_{12} = \mathbf{A}_2^{\prime-1}\mathbf{s}_2^\prime - \mathbf{A}_1^{-1}\mathbf{s}_1
% \end{equation}
\begin{equation}
\mathbf{d}_{12} = \mathbf{A}_{12}^\prime(\mathbf{A}_1^{-1}\mathbf{s}_1-\mathbf{A}_2^{\prime-1}\mathbf{s}_2^\prime)
\end{equation}
which is well-defined modulo the moir\'e unit vector $\mathbf{A}_{12}^\prime$.
Similarly, for the second and third layers,
\begin{equation}
\mathbf{d}_{23} = \mathbf{A}_{23}^\prime(\mathbf{A}_2^{\prime -1}\mathbf{s}_2^\prime-\mathbf{A}_3^{-1}\mathbf{s}_3)
\end{equation}

The relevant parameter in the continuum model of the trilayer is the relative offset of the two moir\'e AA sites, given by
\begin{equation}
\begin{split}
\tilde{\mathbf{d}}=\mathbf{d}_{12}-\mathbf{d}_{23} &=\mathbf{A}_{12}^\prime(\mathbf{A}_1^{-1}\mathbf{s}_1-\mathbf{A}_2^{\prime-1}\mathbf{s}_2^\prime)
-\mathbf{A}_{23}^\prime(\mathbf{A}_2^{-1}\mathbf{s}_2^\prime-\mathbf{A}_3^{-1}\mathbf{s}_3)\\
&=
\mathbf{A}^\prime\left(\frac{1}{\text{p}}(\mathbf{A}_1^{-1}\mathbf{s}_1-\mathbf{A}_2^{\prime-1}\mathbf{s}_2^\prime)
-\frac{1}{\text{q}}(\mathbf{A}_2^{-1}\mathbf{s}_2^\prime-\mathbf{A}_3^{-1}\mathbf{s}_3)\right)
\end{split}
\end{equation}
However, $\mathbf{d}_{12}$ is equivalent up to a shift by $\mathbf{A}_{12}^\prime$, $\mathbf{d}_{12}\cong\mathbf{d}_{12}+\mathbf{A}_{12}^\prime\begin{pmatrix}n\\m\end{pmatrix}$, where $n,m,$ are integers.  
This implies 
\begin{equation}
\tilde{\mathbf{d}} \cong \tilde{\mathbf{d}} + \frac{1}{\text{p}}\mathbf{A}^\prime\begin{pmatrix}n\\m\end{pmatrix} + \frac{1}{\text{q}}\mathbf{A}^\prime\begin{pmatrix}n'\\m'\end{pmatrix}
\end{equation}
Thus, the quantity $\mathbf{d}\equiv|pq|\tilde{\mathbf{d}}$
\begin{equation}
 \mathbf{d}  \cong \mathbf{d} + \mathbf{A}^\prime\begin{pmatrix}qn+pn^\prime\\qm+pm^\prime\end{pmatrix} \cong \mathbf{d}+\mathbf{A}^\prime\begin{pmatrix}n\\m\end{pmatrix}
\end{equation}
is well-defined modulo the commensurate unit cell $\mathbf{A}^\prime$.
In the last step, we have used the fact that since $\text{p}$ and $\text{q}$ are coprime, any integer can be written as an integer multiple $n\text{p}+n^\prime \text{q}$.
Putting everything together, we have the expression 
\begin{equation}
\begin{split}
\mathbf{d}(\mathbf{s}_1,\mathbf{s}_2,\mathbf{s}_3)  &= 
\mathbf{A}^\prime(q\mathbf{A}_1^{-1}\mathbf{s}_1-(p+q)\mathbf{A}_2^{-1}\mathbf{s}_2+p\mathbf{A}_3^{-1}\mathbf{s}_3)
\end{split}
\end{equation}
as quoted in the main text.

\end{document}